\documentclass[aps,pra,showpacs]{revtex4}
\usepackage{graphicx,amsmath,amsfonts,amssymb}

\topmargin by -\headsep \textheight 8.9in \oddsidemargin 0pt
\evensidemargin \oddsidemargin \marginparwidth 0.5in \textwidth
6.5in

\begin{document}

\title{Solitons in quasi one-dimensional Bose-Einstein condensates with
competing dipolar and local interactions}
\author{J. Cuevas$^{1}$, Boris A. Malomed$^{2}$, P. G. Kevrekidis$^{3}$, and
D. J. Frantzeskakis$^{4}$}
\affiliation{$^{1}$Grupo de F\'{\i}sica No Lineal, Departamento de \'{F}isica Aplicada I,
Escuela Universitaria Politecnica, C/ Virgen de \'{A}frica, 7, 41011
Sevilla, Spain \\
$^{2}$Department of Physical Electronics, School of Electrical Engineering,
Faculty of Engineering, Tel Aviv University, Tel Aviv 69978, Israel \\
$3$Department of Mathematics and Statistics, University of Massachusetts,
Amherst, MA 01003-4515, USA \\
$^{4}$Department of Physics, University of Athens, Panepistimiopolis,
Zografos, Athens, 15784, Greece}

\begin{abstract}
We study families of one-dimensional matter-wave bright solitons supported
by the competition of contact and dipole-dipole (DD) interactions of
opposite signs. Soliton families are found, and their stability is
investigated in the free space, and in the presence of an optical lattice
(OL). Free-space solitons may exist with an arbitrarily weak local
attraction if the strength of the DD repulsion is fixed. In the case of the
DD attraction, solitons do not exist beyond a maximum value of the
local-repulsion strength. In the system which includes the OL, a stability
region for \textit{subfundamental solitons} (SFSs) is found in the second
finite bandgap. For the existence of gap solitons (GSs) under the attractive
DD interaction, the contact repulsion must be strong enough. In \ the
opposite case of the DD repulsion, GSs exist if the contact attraction is
not too strong. Collisions between solitons in the free space are studied
too. In the case of the local attraction, they merge or pass through each
other at small and large velocities, respectively. In the presence of the
local repulsion, slowly moving solitons bounce from each other.
\end{abstract}

\date{\today}
\maketitle

\section{Introduction}

Stable localized matter-wave structures in Bose-Einstein condensates (BECs)
are supported by the interplay between the intrinsic nonlinearity, which is
induced by collisions between atoms, quantum pressure, which originates from
the kinetic energy of atoms, and external potentials \cite{review}. This
mechanism has made it possible to create bright solitons in condensates of $%
^{7}$Li and $^{85}$Rb atoms confined in cigar-shaped traps \cite{Li-Rb},
where the inter-atomic interactions are made attractive by means of the
Feshbach resonance (FR) \cite{FR}. In the condensate of $^{87}$Rb atoms with
repulsive interactions, the introduction of an optical-lattice (OL)
potential gives rise to gap solitons (GSs), as demonstrated experimentally
in Ref. \cite{Markus}, see also review \cite{Morsch}. Dark solitons have
been created too, by means of various techniques, in the self-repulsive $%
^{87}$Rb condensate \cite{dark}. In terms of the theoretical description,
the limit case of a very deep OL may be mapped, by means of the
tightly-binding approximation, into a discrete nonlinear Schr\"{o}dinger
equation and, accordingly, GSs are mapped into \textit{staggered} discrete
solitons \cite{deep}.

New possibilities for the formation of matter-wave solitons are suggested by
the presence of long-range interactions in dipolar condensates, which may be
composed of magnetically polarized $^{52}$Cr atoms \cite{Cr}, dipolar
molecules \cite{hetmol}, or atoms in which electric moments are induced by a
strong external field \cite{dc}. Solitons supported by the dipole-dipole
(DD) interactions were predicted in two-dimensional (2D) settings. In the
isotropic configuration, with moments fixed perpendicular to the plane, the
natural DD interaction gives rise to repulsion, which can support
delocalized states in the form of vortex lattices \cite{Pu,Lashkin}. In
principle, the sign of the DD interaction in this configuration may be
reversed by means of rapid rotation of the dipoles \cite{reversal},
suggesting a possibility to create isotropic solitons \cite{Pedri05}, as
well as solitary vortices \cite{Lashkin,Tikh2}. On the other hand, stable
anisotropic solitons have been predicted assuming the natural DD interaction
between dipoles with a fixed in-plane polarization \cite{Tikhonenkov08}. In
addition to these results pertaining to the BEC context, it is relevant to
mention that stable vortex rings were predicted in an optical model with the
nonlocal thermal nonlinearity \cite{Krolik}, and , elliptically shaped
spatial solitons were created in such media experimentally \cite{Moti1}.

Although one-dimensional (1D) configurations may be simpler than their 2D
counterparts, 1D matter-wave bright solitons were not yet studied in detail
in models of dipolar condensates, except for discrete solitons of the
unstaggered type, which were recently predicted in the condensate trapped in
a deep OL \cite{Belgrade}. In those works, both attractive and repulsive
signs of the onsite (contact) nonlinearity and long-range DD interactions
between sites of the respective lattice were considered. The objective of
the present work is the theoretical study of various types of bright
solitons possible in the continuum BEC\ model featuring the competition
between the contact and DD interactions. The effective strength of the DD
interactions in the 1D geometry can be controlled by adjusting the
orientation of the dipoles with respect to the axis of the linear trap,
while the strength of the contact interactions may be effectively tuned by
means of the FR technique, as shown in the condensate of $^{52}$Cr atoms
\cite{recent}. The model is formulated both in the free space (i.e., in the
absence of external potentials) and in the presence of the OL potential,
which opens additional possibilities for the creation of stable localized
states, including GSs (note that staggered discrete solitons, which, as
mentioned above, correspond to GSs in the deep-OL\ limit, were not
considered in Ref. \cite{Belgrade}). In the framework of the
Gross-Pitaevskii equation (GPE) with the ordinary local nonlinear term, the
concept of GSs was elaborated in detail, see Refs. \cite{GS0}-\cite{Louis}
and review \cite{Morsch}. However, to the best of our knowledge, it was not
yet extended to the new physically relevant case, when the OL potential acts
together with the long-range DD interaction -- a situation that we address \
below.

The paper is organized as follows. The model is formulated in Section II,
which, in Section III, is followed by the consideration of solitons
supported by competing nonlinearities\ -- attractive local/repulsive DD, or
vice versa -- in the free space. The model which combines the competing
nonlinearities of both types and the OL is considered in Section IV. In that
case, we report results for regular solitons in the semi-infinite gap, and
for GSs in the two lowest finite bandgaps. In Section V, we deal with
collisions between moving solitons in the absence of the OL. The paper is
concluded by Section VI.

\section{The model}

\begin{figure}[tbp]
\begin{center}
\includegraphics[width=.5\textwidth]{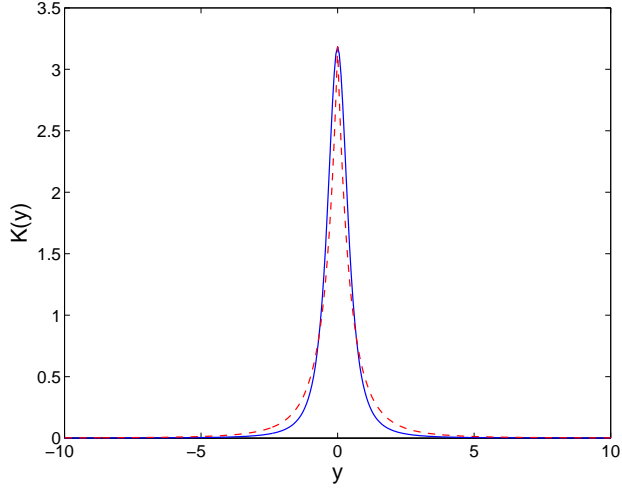}
\end{center}
\caption{(Color online) The solid and dashed lines show the cut-off (CO)
kernel, with $y_{c}=\protect\pi ^{-1/2}$, and its single-mode approximation
(SMA) counterpart, which are based on Eqs. (\protect\ref{eq:kernel0}) and (%
\protect\ref{eq:kernel1}), respectively. }
\label{fig:dipole}
\end{figure}

Our aim is to construct soliton states within the framework of the 1D GPE
for the mean-field wave function, $\psi (x,t)$. The scaled equation includes
the OL potential, $V(x)=\epsilon \sin ^{2}x$ (where $\epsilon $ is the
strength of the OL, while its period is normalized to be $\pi $), the local
nonlinear term with the respective coefficient, $g_{c}$, and its nonlocal
counterpart, with coefficient $g_{d}$, which accounts for the DD
interactions:
\begin{equation}
i\psi _{t}=-\frac{1}{2}\psi _{xx}+\left[ \epsilon \sin ^{2}x+g_{c}|\psi
|^{2}+g_{d}\int_{-\infty }^{+\infty }K(x-x^{\prime })|\psi (x^{\prime })|^{2}%
\mathrm{d}x^{\prime }\right] \psi ,  \label{GPE}
\end{equation}%
where subscripts denote partial derivatives. The kernel $K$ of the DD term
was taken in two different ways, so as to avoid the divergence at $%
x=x^{\prime }$. The first version relies on an explicit cut-off (\textit{CO
kernel}),
\begin{equation}
K(y)=y_{c}^{3}(y^{2}+y_{c}^{2})^{-3/2},  \label{eq:kernel0}
\end{equation}%
where $y_{c}$ is a constant. The other choice makes use of the regularized
expression deduced in Ref. \cite{Sinha} by means of the single-mode
approximation (SMA), i.e., the \textit{SMA kernel}:
\begin{equation}
K(y)=\frac{10}{\pi }\left[ (1+2y^{2})\exp (y^{2})\mathrm{erfc}(|y|)-2\frac{%
|y|}{\sqrt{\pi }}\right] ,  \label{eq:kernel1}
\end{equation}%
with $\mathrm{erfc}(y)$ being the standard complimentary error function. The
two kernels are compared in Fig. \ref{fig:dipole}. The most significant
difference between them is that the SMA version features a cusp at $y=0$,
whereas the CO kernel has a smooth maximum. We have concluded that the
choice of $y_{c}=\pi ^{-1/2}$ in CO expression (\ref{eq:kernel0}), which
makes areas beneath both curves equal, provides for the best proximity of
the corresponding results to what has been found using the SMA
approximation. Below, we report results obtained with the SMA kernel, as its
CO counterpart yields virtually identical findings.

We fix the normalizations in Eq. (\ref{GPE}) by setting $g_{d}=\pm 1$, and
then vary $g_{c}$. In the the case of the $^{52}$Cr condensate, a
characteristic value of the relative strength of the DD and contact
interactions, which may be estimated as $\left\vert g_{d}/g_{c}\right\vert $%
, is $\simeq 0.15$ \cite{recent}; this value may be altered in broad limits
by means of the FR technique. The interactions are repulsive or attractive
for $g_{d},g_{c}>0$ and $g_{d},g_{c}<0$, respectively. We will focus on the
case of \textit{competing} interactions, with $g_{c}g_{d}<0$, which is the
most interesting one; in the case when both nonlinearities have the same
sign, results turn out to be very similar to those reported previously in
the local model. It is relevant to mention that, in terms of discrete
systems, the competition of on-site (local) and inter-site (short-range
nonlocal) interactions in 1D and 2D Salerno lattices were considered in
Refs. \cite{Zaragoza} and \cite{Zaragoza2}, respectively. A number of stable
discrete-soliton states which are impossible in the absence of the
competition were reported in those works, including cuspons and peakons in
1D, and vortex breathers in 2D.

Stationary solutions to Eq. (\ref{GPE}) with chemical potential $\mu $ are
sought as $\psi (x,t)=\Psi (x)\exp (-i\mu t)$. To construct such solutions,
we discretize the resulting equation for $\Psi (x)$ by means of a
finite-difference scheme, which leads to a set of coupled algebraic
equations,
\begin{gather}
\mu \Psi _{n}=-\frac{1}{2(\Delta x)^{2}}(\Psi _{n+1}+\Psi _{n-1}-2\Psi _{n})
\notag \\
+\left( \epsilon \sin ^{2}x_{n}+g_{c}\Psi _{n}^{2}+g_{d}\Delta
x\sum_{m}K_{n-m}\Psi _{m}^{2}\right) \Psi _{n},  \label{GPEdiscr}
\end{gather}%
with $x_{n}=n\Delta x$, and $K_{n-m}=K(|n-m|\Delta x)$. Solutions to Eq. (%
\ref{GPEdiscr}) are sought by dint of the Newton-Raphson scheme. Results
were obtained with a reasonable accuracy by choosing $\Delta x=\pi /40$. To
present soliton families, we will use norm $N$ and width $W$ of the soliton,
\begin{equation}
N=\int_{-\infty }^{\infty }|\Psi |^{2}\ \mathrm{d}x,~W=\sqrt{%
N^{-1}\int_{-\infty }^{\infty }x^{2}|\Psi |^{2}\ \mathrm{d}x~}.  \label{W}
\end{equation}%
These definitions were adapted to the finite-difference form of the model as
well.

Stability of the solutions was analyzed in the standard way (see, e.g., Ref.
\cite{Pelinovsky}), by considering a perturbation in the form of $\delta
\psi (x,t)=\exp (-i\mu t)[P(x)\exp (-i\lambda t)+Q^{\ast }(x)\exp (i\lambda
^{\ast }t)]$ (where $\ast $ stands for complex conjugate). The linearized
equations for the perturbation eigenmodes (i.e., the Bogoliubov - de Gennes
equations) are written as
\begin{equation}
\lambda \left(
\begin{array}{c}
P(x) \\
Q^{\ast }(x)%
\end{array}%
\right) =\left(
\begin{array}{cc}
\hat{L}_{1} & \hat{L}_{2} \\
-\hat{L}_{2}^{\ast } & -\hat{L}_{1}^{\ast }%
\end{array}%
\right) \left(
\begin{array}{c}
P(x) \\
Q^{\ast }(x)%
\end{array}%
\right) ,  \label{BdG}
\end{equation}%
where we define
\begin{eqnarray}
\hat{L}_{1}\equiv - &&\mu -\frac{1}{2}\partial _{x}^{2}+V(x)+2g_{c}|\psi
(x)|^{2} \\
&+&g_{d}\int_{-\infty }^{+\infty }\mathrm{d}x^{\prime }K\left( x-x^{\prime
}\right) \left[ \psi ^{\ast }(x^{\prime })\psi (x)+|\psi (x^{\prime })|^{2}%
\right] ,
\end{eqnarray}%
\begin{equation}
\hat{L}_{2}\equiv g_{c}\psi ^{2}(x)+g_{d}\int_{-\infty }^{+\infty }\mathrm{d}%
x^{\prime }K(x-x^{\prime })\psi (x^{\prime })\psi (x).
\end{equation}%
The instability sets in when there emerges an eigenvalue with $\mathrm{Im}%
(\lambda )\neq 0$. Stability eigenvalues were obtained from a numerical
solution of Eq. (\ref{BdG}), by means of a standard eigenvalue solver from
Fortran-based software package LAPACK \cite{LAPACK}.

\section{Solitons in the absence of the optical lattice}

We start the presentation of results by considering the model with competing
nonlinearities in the free space, i.e., with $\epsilon =0$ in Eq. (\ref{GPE}%
). This implies that solitons may exist with $\mu <0$.

\subsection{Attractive local and repulsive nonlocal interactions}

In accordance with what was said above, we first fix $g_{d}=1$ (the
repulsive DD interaction), and vary negative $g_{c}$ (local attraction) and
negative $\mu $. In this case, solitons exist for every $g_{c}<0$; however,
due to discretization, the numerical solution of Eq. (\ref{GPEdiscr}), with
the above-mentioned choice of $\Delta x=\pi /40$, yields solitons in the
region of $\left\vert g_{c}\right\vert >0.25$. In fact, although very narrow
solitons indeed exist at arbitrarily small values of $-g_{c}$, the soliton's
width becomes comparable to or smaller than this value of $\Delta x$ at $%
\left\vert g_{c}\right\vert \lesssim 0.9$, i.e., a better numerical accuracy
is required to produce accurate soliton solutions in this range. Note that,
for very narrow solitons with amplitude $\Psi _{0}$, the nonlinear part of
Eq. (\ref{GPEdiscr}), with $g_{d}\equiv 1$, takes the form of $\left(
g_{c}+K_{0}\Delta x\right) \Psi _{0}^{3}$. The existence of the soliton
demands a negative coefficient in this expression, i.e.,
\begin{equation}
\left\vert g_{c}\right\vert >(10/\pi )\Delta x,  \label{Deltax}
\end{equation}%
where it was taken into regard that $K_{0}=10/\pi $, as per Eq. (\ref%
{eq:kernel1}). In particular, for $\Delta x=\pi /40$, condition (\ref{Deltax}%
) amounts to $\left\vert g_{c}\right\vert >0.25$, as said above.

\begin{figure}[tbp]
\begin{center}
\begin{tabular}{cc}
\includegraphics[width=.5\textwidth]{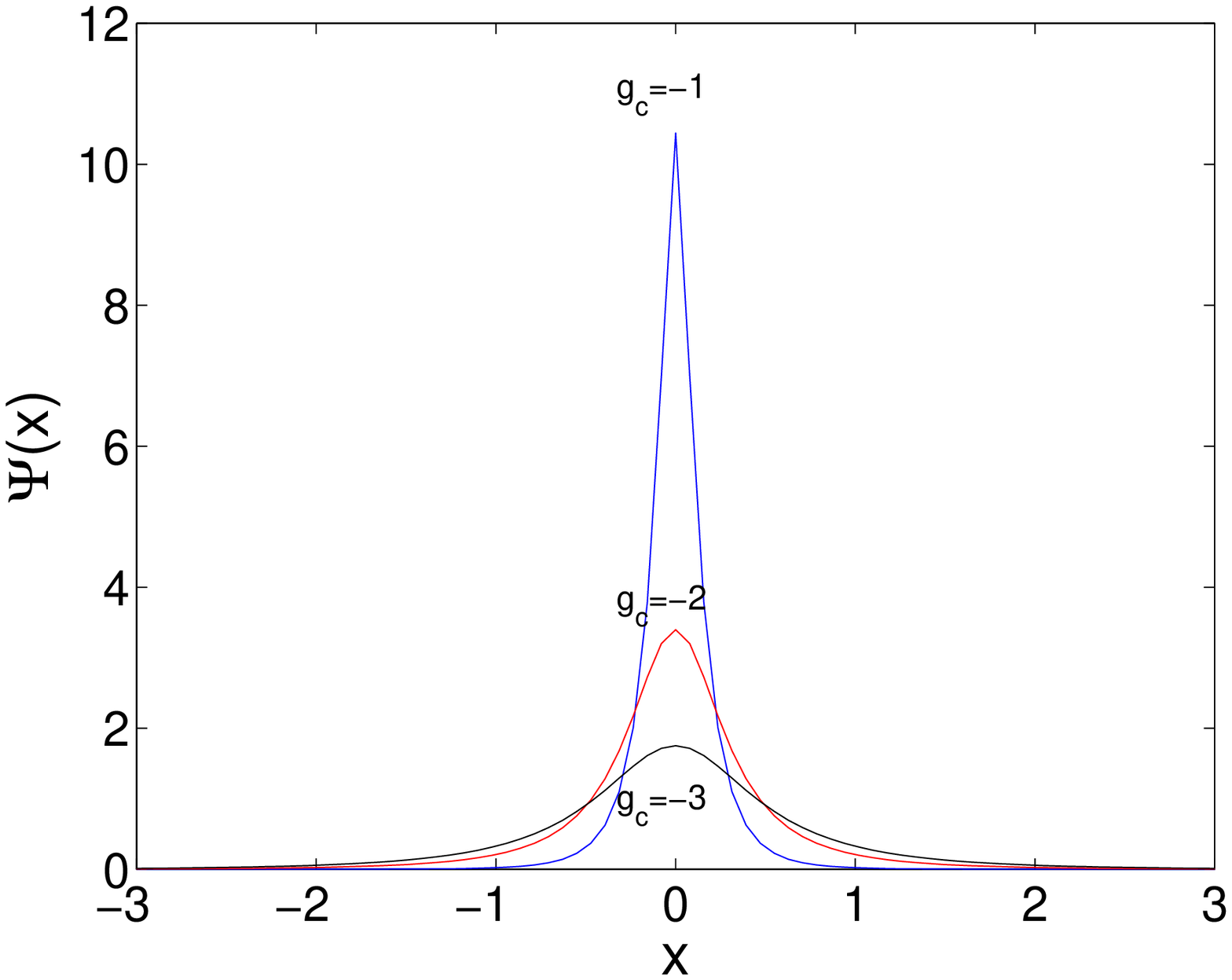} & \includegraphics[width=.5\textwidth]{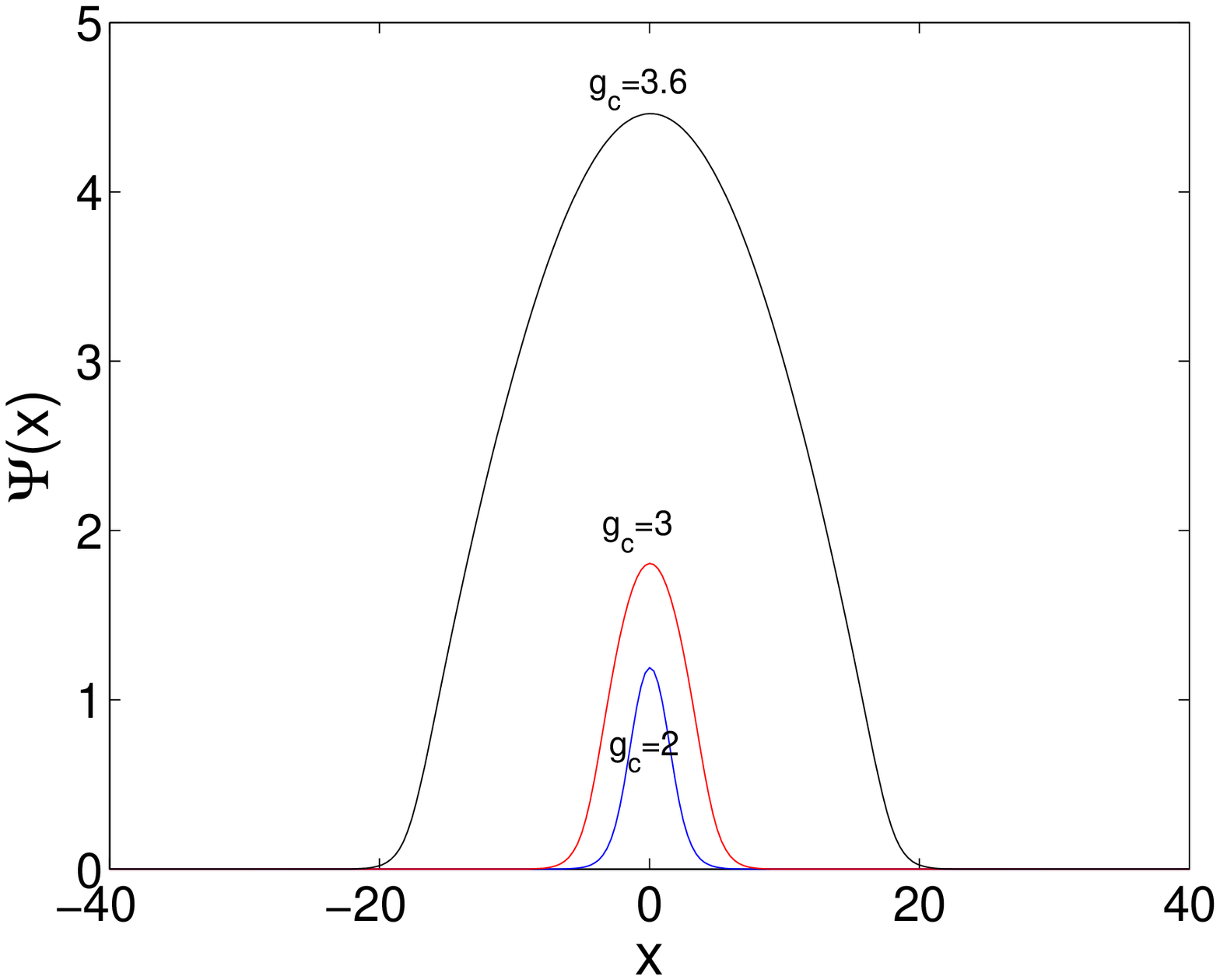} \\
(a) & (b)%
\end{tabular}%
\end{center}
\caption{(Color online) (a) Profiles of stable solitons in the model with
attractive local and repulsive nonlocal interactions. The solitons were
found in the numerical form with resolution $\Delta x=\protect\pi /40,$
which yields accurate results for $\left\vert g_{c}\right\vert >0.9$. The
peak gets sharper as $|g_{c}|$ decreases. (b) Profiles of solitons in the
model with repulsive local and attractive nonlocal interactions, obtained
with $\Delta x=\protect\pi /10$. The compacton-like solution corresponding
to $\left\vert g_{c}\right\vert =3.6$ is close to the critical point, beyond
which (at larger $\left\vert g_{c}\right\vert $) solitons do not exist,
irrespective of $\Delta x$. In both panels, all solitons pertain to $\protect%
\mu =-1$.}
\label{fig:profiles}
\end{figure}

\begin{figure}[tbp]
\begin{center}
\begin{tabular}{cc}
\includegraphics[width=.5\textwidth]{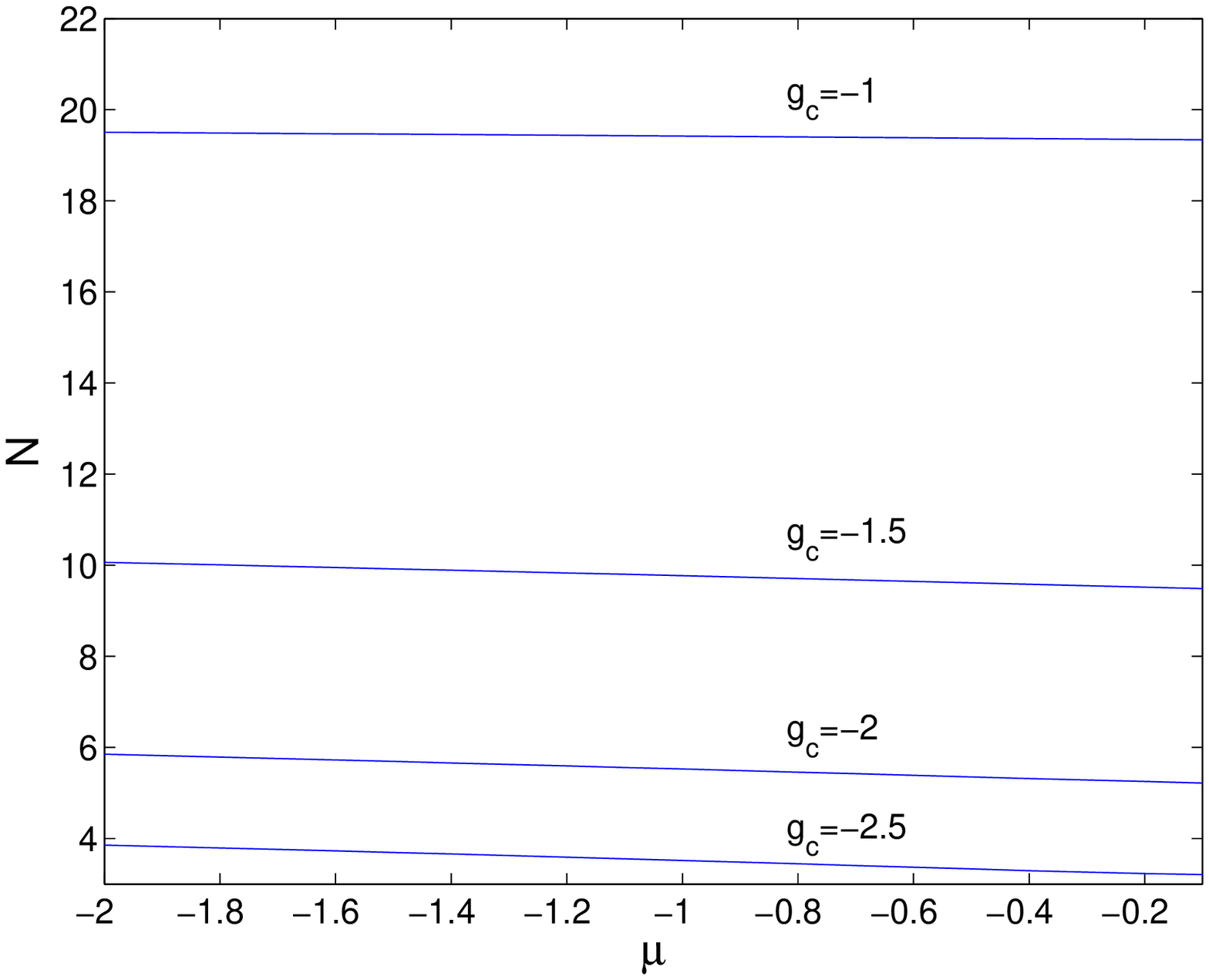} & \includegraphics[width=.5\textwidth]{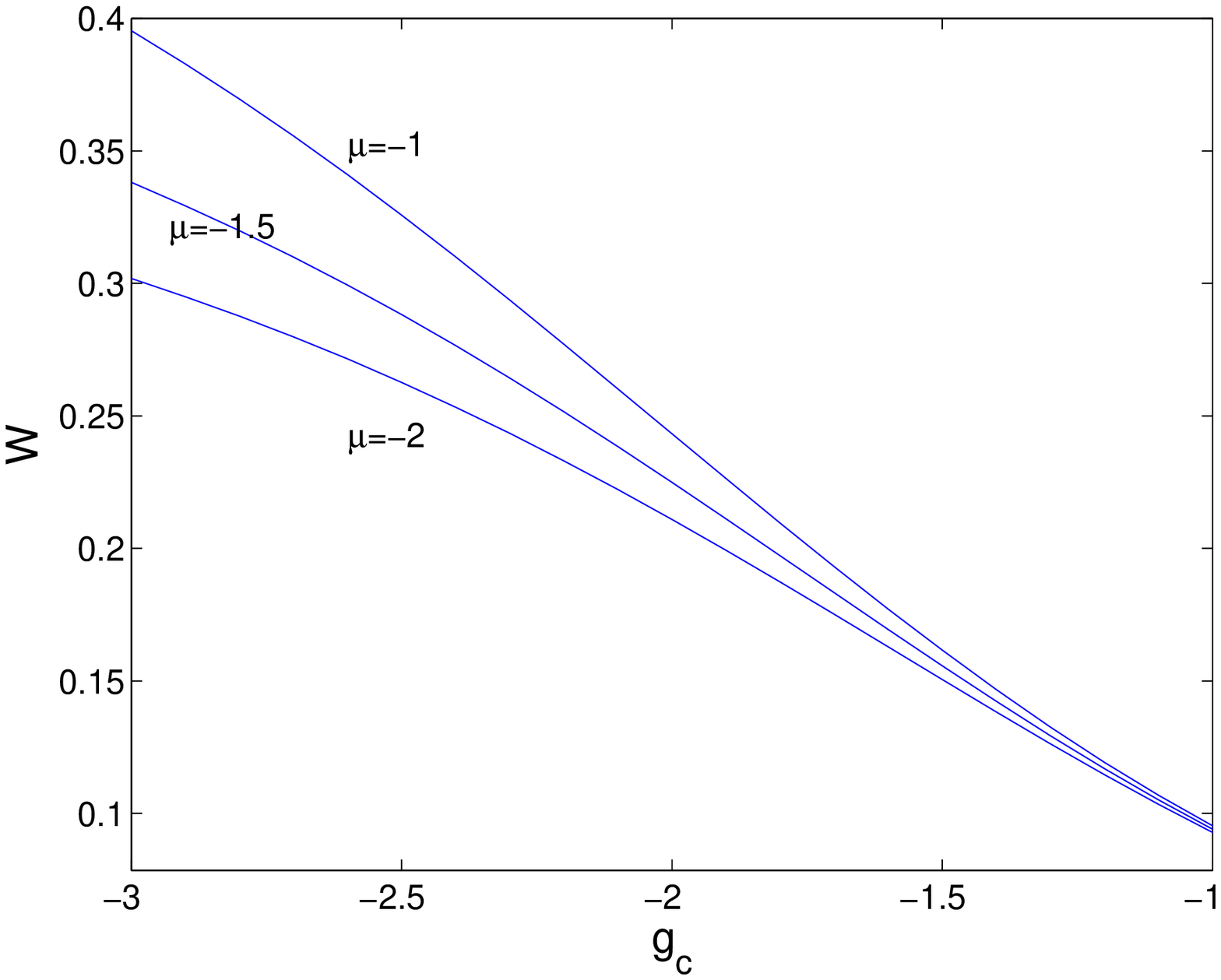} \\
(a) & (b)%
\end{tabular}%
\end{center}
\caption{(Color online) (a) The norm of solitons in the free space, in the
case of attractive local and repulsive nonlocal interactions, versus the
chemical potential. (b) The width of the solitons, defined as per Eq. (%
\protect\ref{W}), versus the strength of the local attraction. In panel (b),
the plots are shown for $W>\Delta x=\protect\pi /40$: recall that obtaining
numerical results for smaller $W$ requires using smaller $\Delta x$. }
\label{fig:normar}
\end{figure}

Figure \ref{fig:profiles}(a) shows typical examples of the solitons in the
present case, while Fig. \ref{fig:normar} represents soliton families in
terms of dependences $N(\mu )$ and $W\left( g_{c}\right) $, for fixed values
of $g_{c}$ and $\mu $, respectively; note that plots in panel (b) of the
latter figure are cut at $g_{c}=-1$, as the numerical accuracy is
insufficient to extend them to smaller values of $\left\vert
g_{c}\right\vert $, as explained above. The stability of the solitons was
verified both through the computation of the eigenvalues, using Eq. (\ref%
{BdG}), and by means of direct simulations of the evolution of perturbed
solitons. The well-known Vakhitov--Kolokolov criterion, $dN/d\mu <0$ \cite%
{VK}, also suggests the stability of solitons in this case, although the
negative slope of the $N(\mu )$ curve in Fig. \ref{fig:normar}(a) is very
small.

\subsection{Repulsive local and attractive nonlocal interactions}

Now, we fix $g_{d}=-1$, varying $g_{c}>0$ and $\mu <0$. Contrary to the
previous case, solitons (which are stable) can be readily found for all
values of $g_{c}$ up to $g_{c}\approx 3.6$, the difference between $g_{c}=0$
(zero local interaction, while the DD attraction is present) and $g_{c}>0$
amounting to a gradual increase of the soliton's amplitude and width with $%
g_{c}$. As seen in Fig. \ref{fig:profiles}(b), close to $g_{c}=3.6$ the
soliton develops a compacton-like shape, and solitons cannot be found at $%
g_{c}>\left( g_{c}\right) _{\max }\approx 3.7$.

\begin{figure}[tbp]
\begin{center}
\begin{tabular}{cc}
\includegraphics[width=.5\textwidth]{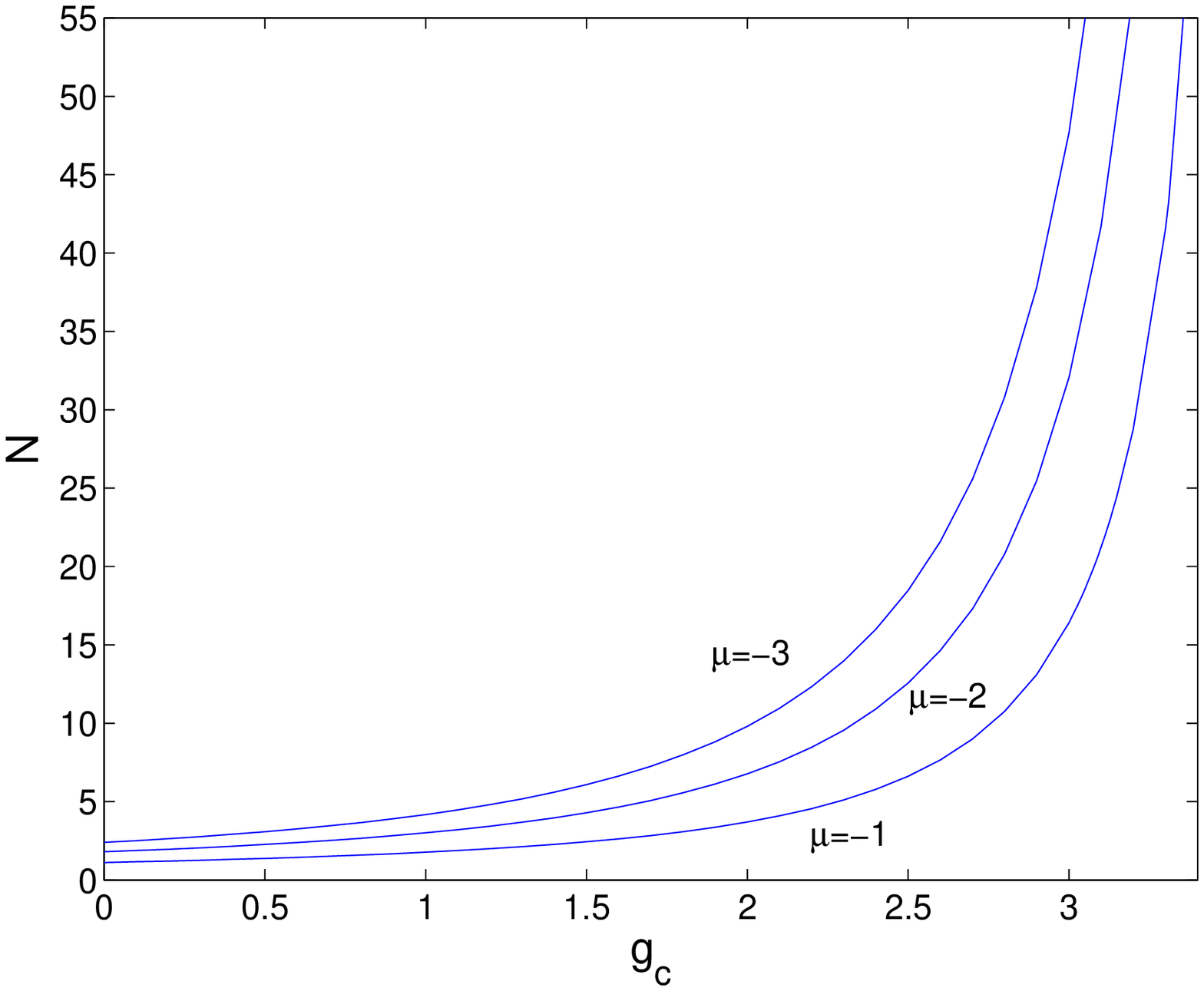} & \includegraphics[width=.5\textwidth]{{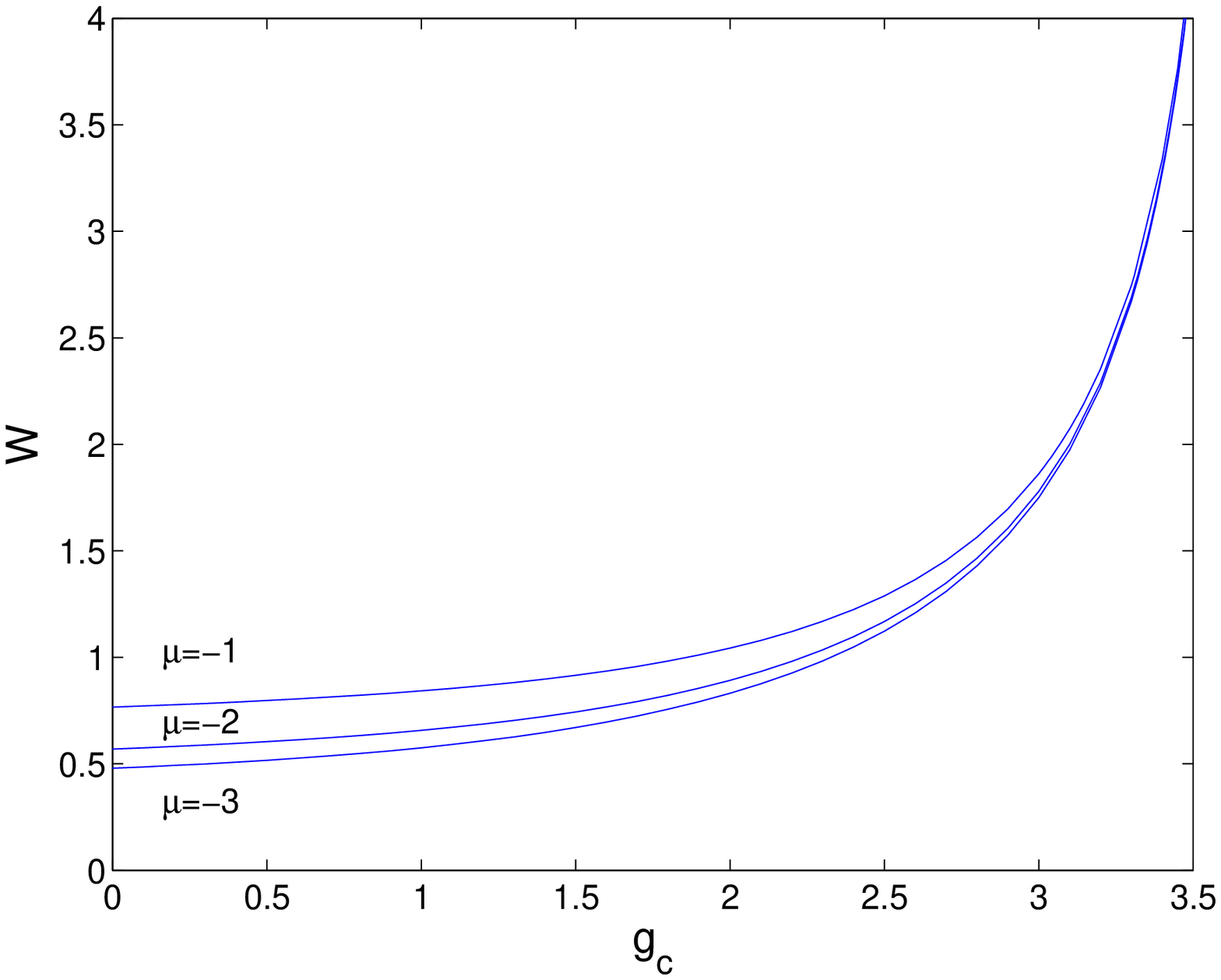}}%
\end{tabular}%
\end{center}
\caption{(Color online) The norm and width of fundamental solitons in the
free space, in the case of attractive nonlocal and repulsive local
interactions, versus the strength of the local repulsion.}
\label{fig:normra}
\end{figure}

The existence of $\left( g_{c}\right) _{\max }$ can be easily explained.
Indeed, in the limit of $\Delta x\rightarrow 0,$ and for a very broad
soliton, the nonlinear part of Eq. (\ref{GPEdiscr}), with $g_{d}\equiv -1$,
takes the approximate form of
\begin{equation}
\left[ g_{c}-\int_{-\infty }^{+\infty }K(y)\mathrm{d}y\right] |\psi
(x)|^{2}\psi (x).  \label{cubic}
\end{equation}%
The necessary condition for the existence of solitons is that the
coefficient in front of $|\psi (x)|^{2}\psi (x)$ in this expression must be
negative [cf. the derivation of Eq. (\ref{Deltax})], i.e.,
\begin{equation}
g_{c}<\left( g_{c}\right) _{\max }\equiv \int_{-\infty }^{+\infty }K(y)%
\mathrm{d}y=20\pi ^{-3/2}\approx 3.59,  \label{max}
\end{equation}%
where expression (\ref{eq:kernel1}) was used to perform the integration. For
finite $\Delta x$, the integral in expression (\ref{max}) is replaced by $%
\sum_{m}K_{m}\Delta x$. In particular, for $\Delta x=\pi /10$, this yields $%
\left( g_{c}\right) _{\max }\approx 3.708$, which is consistent with the
above-mentioned numerical finding. Note also that the vanishing of the
coefficient in front of $|\psi (x)|^{2}\psi (x)$ in expression (\ref{cubic})
at $\left( g_{c}\right) _{\max }-g_{c}\rightarrow 0$ implies that the
soliton's amplitude and norm diverge in this limit, which is corroborated by
the numerical results shown in Fig. \ref{fig:normra}.

\begin{figure}[tbp]
\begin{center}
\begin{tabular}{cc}
\includegraphics[width=.5\textwidth]{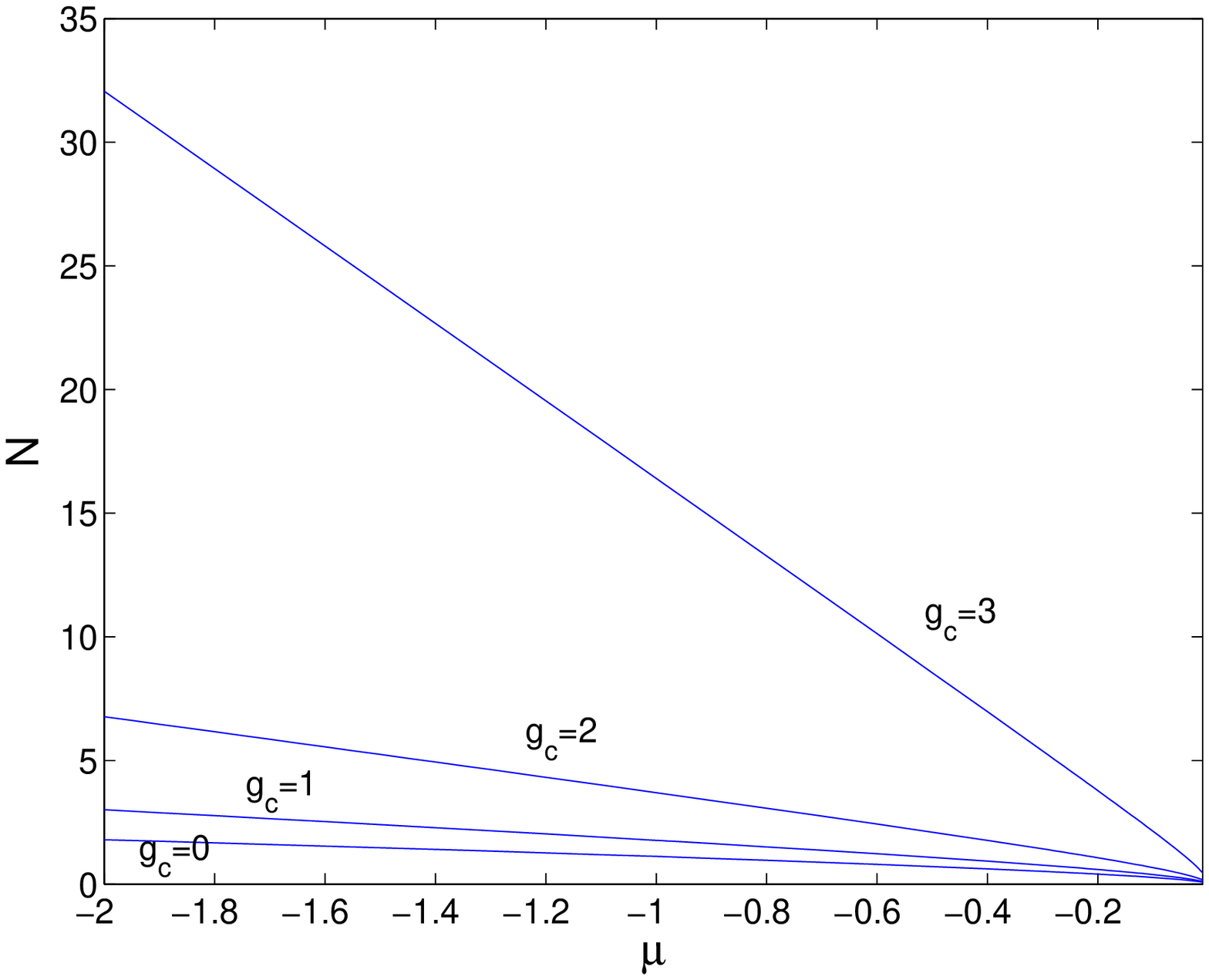} & \includegraphics[width=.5\textwidth]{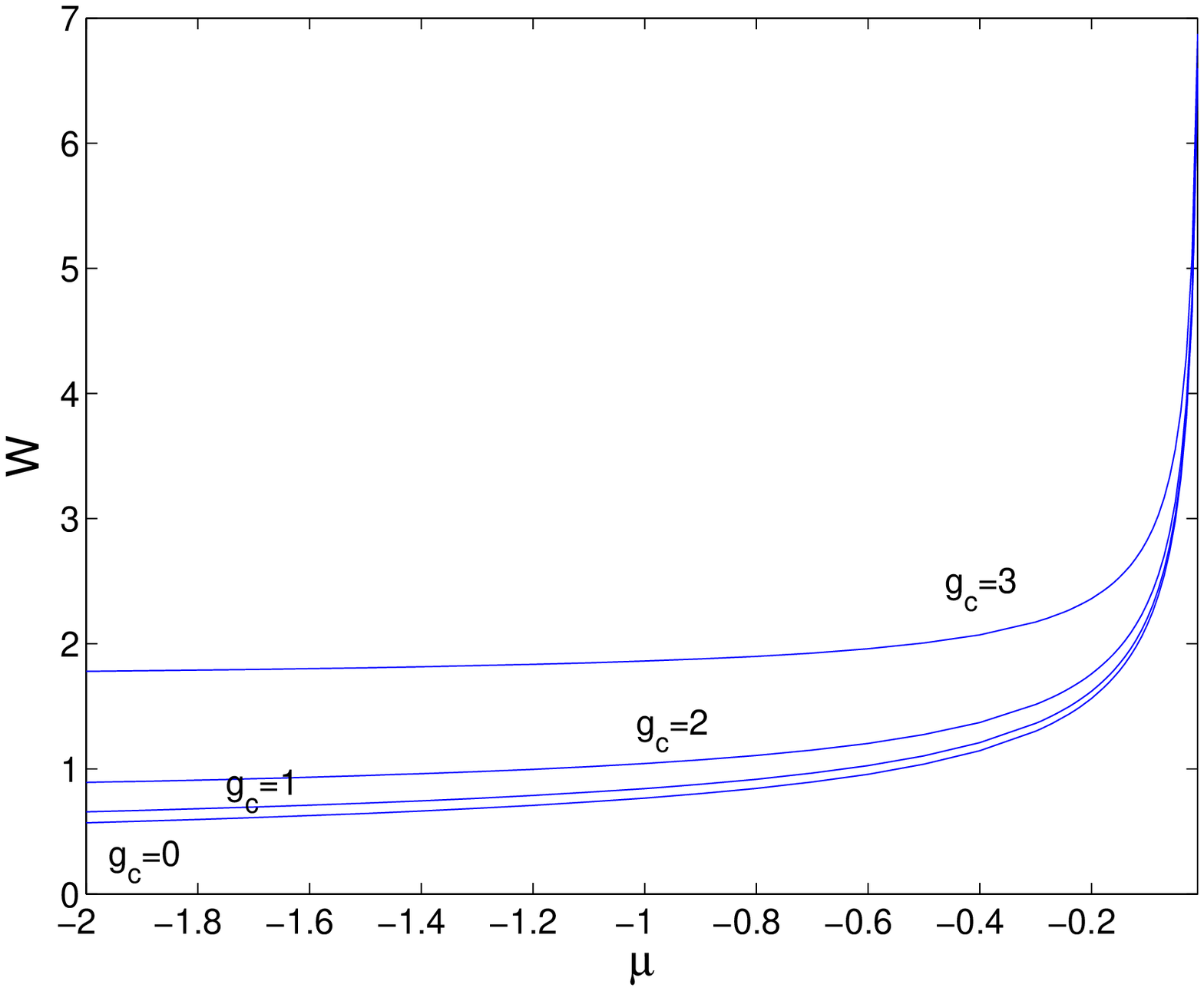} \\
&
\end{tabular}%
\end{center}
\caption{(Color online) The norm and width of fundamental solitons in the
free space, in the case of attractive nonlocal and repulsive local
interactions, versus the chemical potential.}
\label{fig:widthra}
\end{figure}

The dependence of the soliton's norm and width on the chemical potential is
displayed in Fig. \ref{fig:widthra}. Unlike the nearly flat $N(\mu )$
dependences in Fig. \ref{fig:normar}, the present ones clearly satisfy the
VK stability criterion, $dN/d\mu <0$. The full stability of the solitons was
confirmed by the computation of eigenvalues, using Eq. (\ref{BdG}).

\begin{figure}[tbp]
\includegraphics[width=.5\textwidth]{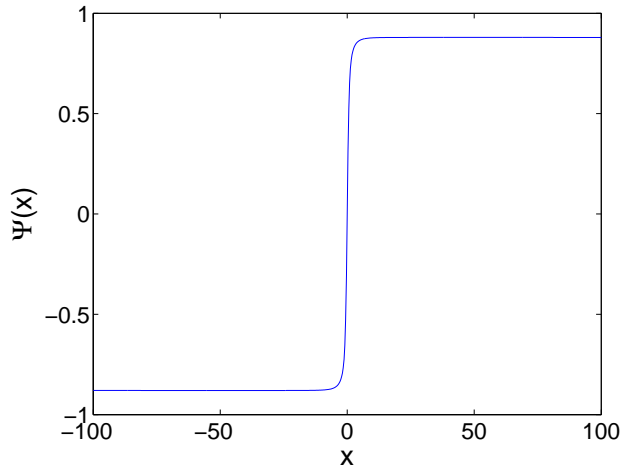}
\caption{(Color online) The profile of a stable dark soliton in the model
with attractive local and repulsive nonlocal interactions, for $g_{c}=5$, $%
g_{d}=-1$, and $\protect\mu =1$. The soliton was found in the numerical form
with resolution $\Delta x=\protect\pi /10$.}
\label{fig:dark}
\end{figure}

At this point, it is also worth to briefly consider dark solitons, which are
known to be stable in BEC with contact repulsive interactions, such as $%
^{87} $Rb condensates \cite{review,dark}. Dark solitons in three-dimensional
dipolar BECs were recently considered in Ref. \cite{ddark}, where it was
shown that, for sufficiently strong repulsive DD interactions and a
sufficiently deep OL in the soliton's nodal plane, dark solitons exist and
are stable. In the present 1D setup, to investigate the existence and
stability of dark solitons, it is first necessary to ensure that the
respective background, namely, constant-amplitude state $\psi =\sqrt{\mu }%
\exp (-i\mu t)$, is modulationally stable. A comprehensive analysis of the
modulational instability (MI) of the background in the context of Eq. (\ref%
{GPE}) can be performed following the lines of Ref. \cite{nMI}. Here we will
briefly consider this issue and provide an example of a stable dark soliton,
assuming $g_{c}+(20/\pi ^{3/2})g_{d}>0$. In this case, the effective
nonlinearity for long-wavelength perturbations (i.e., those with wave
numbers $k\rightarrow 0$) is self-defocusing, i.e., the MI band cannot start
at $k=0$, as it does in the case of the standard nonlinear Schr\"{o}dinger
equation with the self-focusing nonlinearity. Although the MI band may
appear at finite $k$, this is not expected to happen as long as the
background density is small enough, because the maximum MI gain is
proportional to that density. Thus, in this case, a modulationally stable
background may exist, and dark-soliton solutions can be found. As an
example, in Fig. \ref{fig:dark} we show a stable dark soliton (its stability
was verified through the computation of the full spectrum of eigenvalues for
small perturbations), which was found for $\mu =1$ and $g_{c}=5$, $g_{d}=-1$%
. A systematic analysis of the MI of the background and dark-soliton
families in the framework of Eq. (\ref{GPE}) is beyond the scope of this
work, and is deferred to a separate publication.

\section{Solitons in the optical lattice}

In the presence of the OL potential, generic results for regular solitons
and gap solitons (GSs) in the model with the competing interactions can be
adequately represented by fixing the OL strength to $\epsilon =6$, which is
adopted below. GSs have been found in the first and second finite bandgaps
of the OL-induced linear spectrum. For $\epsilon =6$, the two numerically
computed (with $\Delta x=\pi /40$) bandgaps cover, respectively, the
following intervals of the chemical potential:
\begin{equation}
1.61<\mu <4.26~\mathrm{and~}4.63<\mu <6.02.  \label{12}
\end{equation}

\subsection{Attractive local and repulsive nonlocal interactions}

Here, we consider solitons in the case of the competition between local
attraction ($g_{c}<0$) and repulsive DD interactions ($g_{d}=1$), varying $%
\mu $ and $g_{c}$. The self-focusing character of the local interaction
allows the existence of regular solitons in the semi-infinite gap, which is $%
-\infty <\mu <1.5810$ for $\epsilon =6$. The numerical solution of Eq. (\ref%
{GPEdiscr}), with $\Delta x=\pi /40$, yields regular solitons for $%
g_{c}<-0.25$, similar to the case of $\epsilon =0$, see above.

Apart from the solitons in the semi-infinite gap, GSs have been found in
parts of the first and second finite bandgaps, for $g_{c}$ exceeding a
certain critical value, which depends on $\mu $, as shown in Fig. \ref%
{fig:existar}. Multi-humped solitons, which are bound states of fundamental
single-humped solitons, can be found too. Numerical results demonstrate that
the existence range for the multi-humped solitons is slightly broader than
for the fundamental ones. In the second bandgap, single-humped solitons
exist (and are stable) at $\mu >\mu _{\mathrm{cr}}\approx 5.1$. This
boundary value, which corresponds to the vertical dashed line in Fig. \ref%
{fig:existar}, is located above the lower edge of the second bandgap, $\mu
\approx 4.63$, see Eq. (\ref{12}).

\begin{figure}[tbp]
\begin{center}
\includegraphics[width=.5\textwidth]{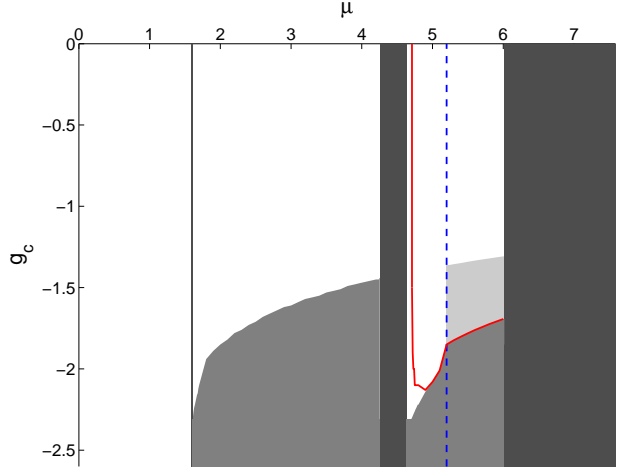}
\end{center}
\caption{(Color online) Unshaded and light-gray-shaded areas represent
existence regions for solitons in the semi-infinite and two lowest finite
gaps, in the model combining the optical lattice with competing attractive
local and repulsive nonlocal interactions. Dark-gray shading covers Bloch
bands, where solitons do not exist. In the gray-shaded parts of the finite
bandgaps (intermediate between light and dark gray), the solitons do not
exist either. In the white part of the second bandgap on the left side of
the vertical dashed line, only subfundamental solitons (SFSs) exist. They
are stable between the red (solid) line and the gray area. On the right side
of the dashed line, stable fundamental solitons exist in the white region,
while unstable SFSs exist both in the white and light-gray areas.}
\label{fig:existar}
\end{figure}

Another species of stable GSs was found in a part of the second
finite bandgap, in the form of subfundamental solitons (SFSs). These
are antisymmetric modes which are squeezed, essentially, into a
single cell of the OL. The norm of the SFS is smaller than that of
the fundamental GS, if the latter one can be found at the same value
of $\mu $ (hence the name of ``subfundamental" \cite{Thawatchai}).
In a narrow interval adjacent to the lower edge of the second
bandgap, $4.63\leq \mu <4.71$, the SFSs are stable (see Fig.
\ref{fig:existar}), whereas above this interval, they undergo a
Hamiltonian-Hopf bifurcation, being unstable for $\left\vert
g_{c}\right\vert $ smaller than a certain critical value. In the
region of $\mu >\mu _{\mathrm{cr}}\approx 5.1$, SFSs are unstable in
their entire existence region. In direct simulations, the
destabilized SFSs spontaneously transform themselves into stable
fundamental solitons belonging to the first (rather than second)
bandgap; a similar scenario of
the instability development of SFSs was found in the local model \cite%
{Thawatchai}.

\begin{figure}[tbp]
\begin{center}
\includegraphics[width=.5\textwidth]{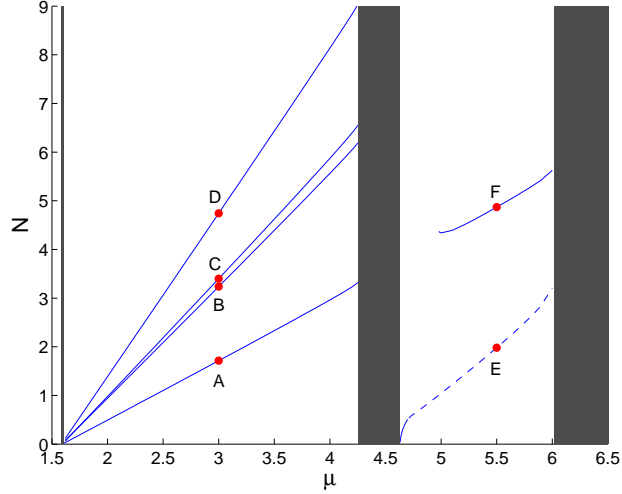}
\end{center}
\caption{(Color online) $N(\protect\mu )$ curves for gap solitons in the
model including the optical lattice, local attraction, and nonlocal
repulsion. The strength of the local attraction is fixed to $g_{c}=-1$.}
\label{fig:normgapar}
\end{figure}

\begin{figure}[tbp]
\begin{center}
\begin{tabular}{cccc}
\includegraphics[width=.2\textwidth]{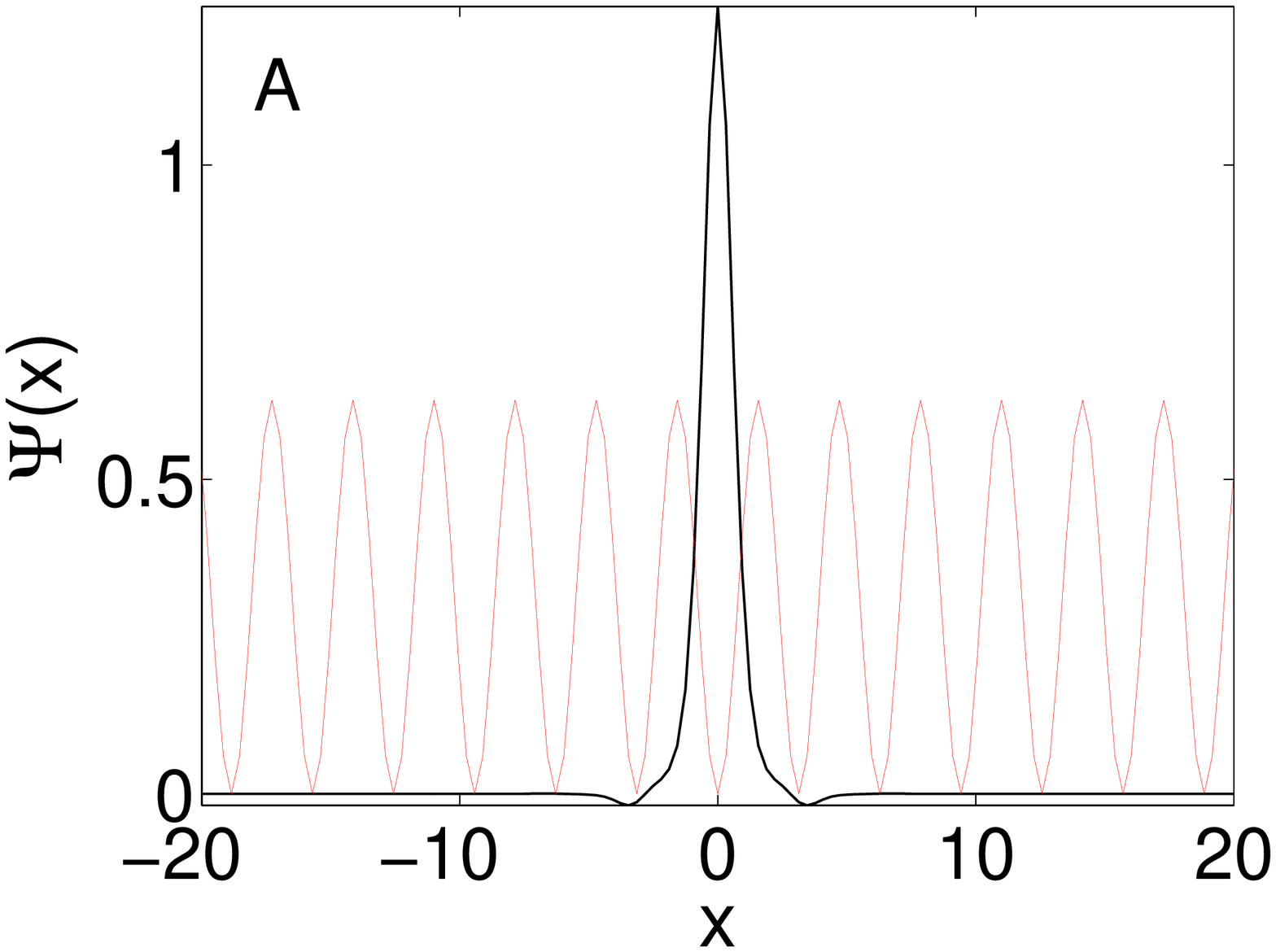} & \includegraphics[width=.2\textwidth]{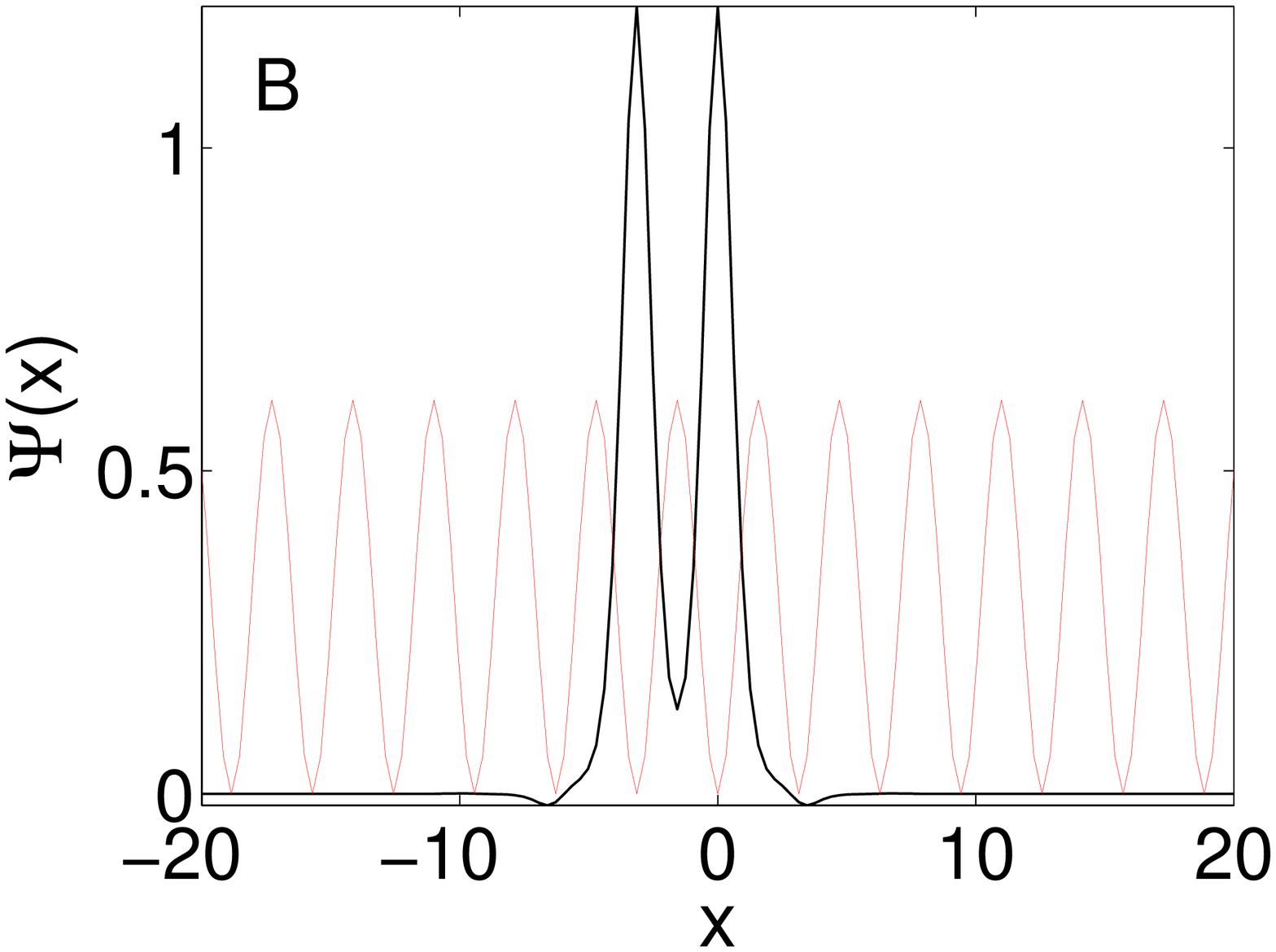} & \includegraphics[width=.2\textwidth]{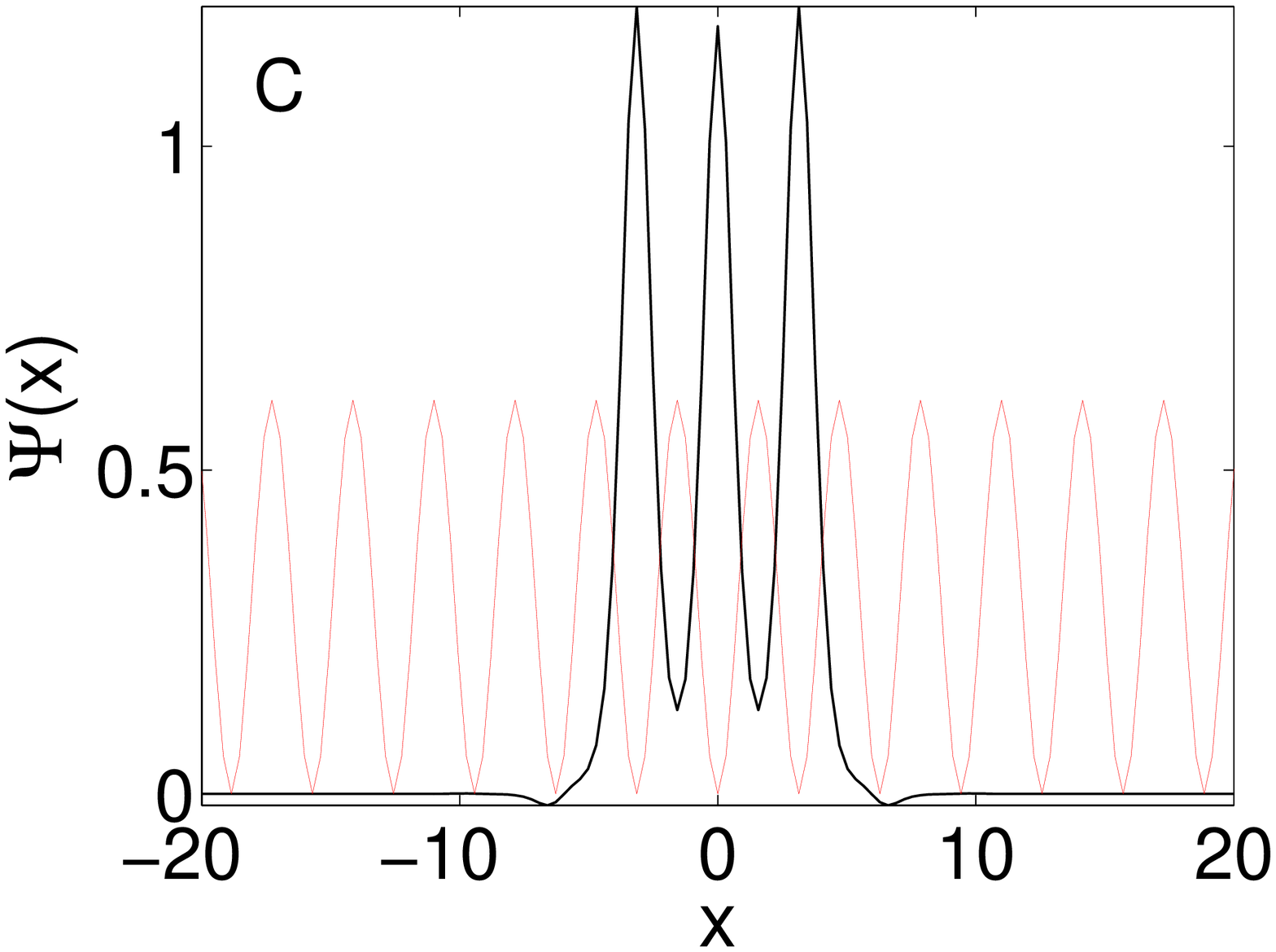} & 
\includegraphics[width=.2\textwidth]{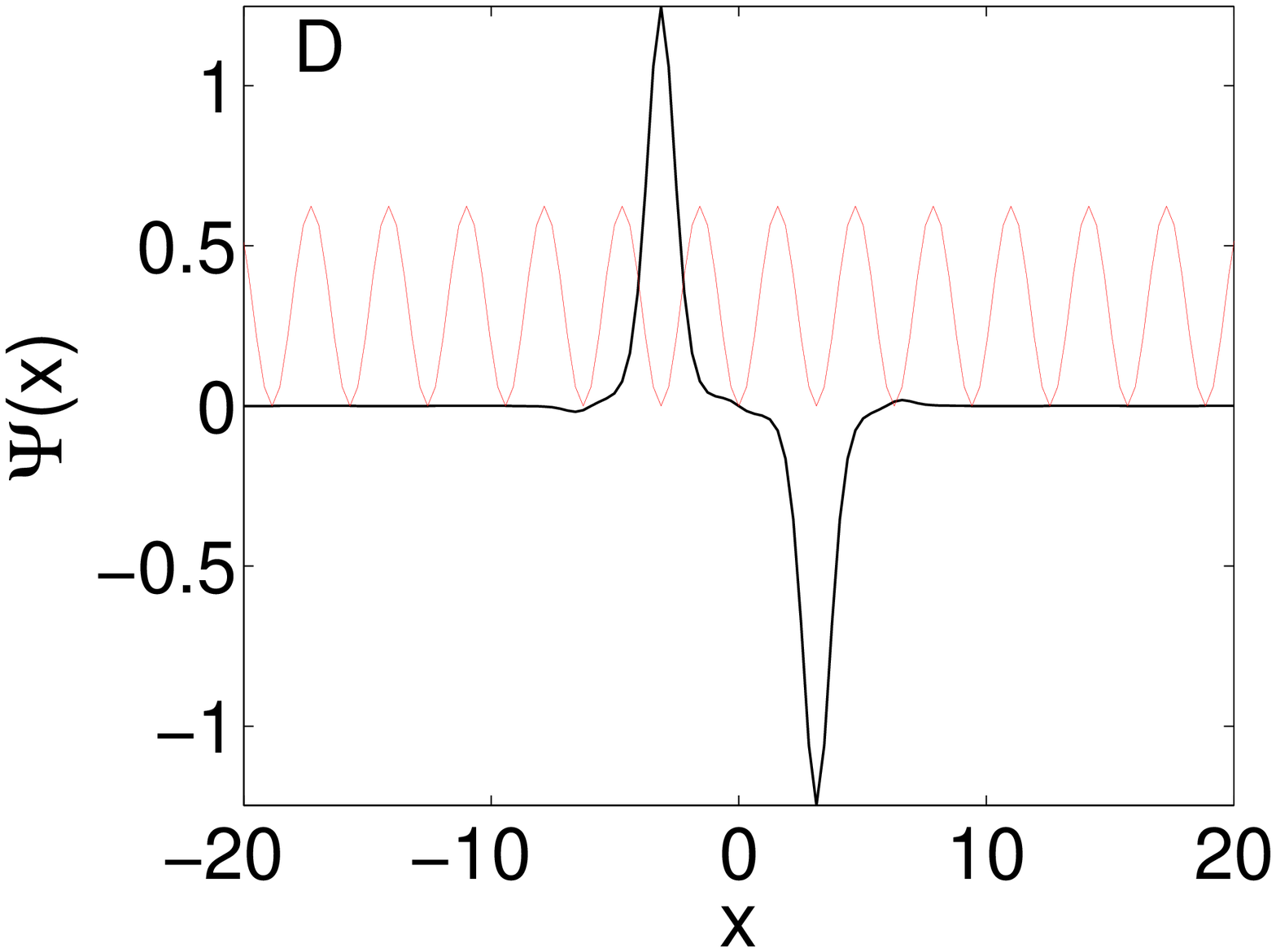} \\
\includegraphics[width=.2\textwidth]{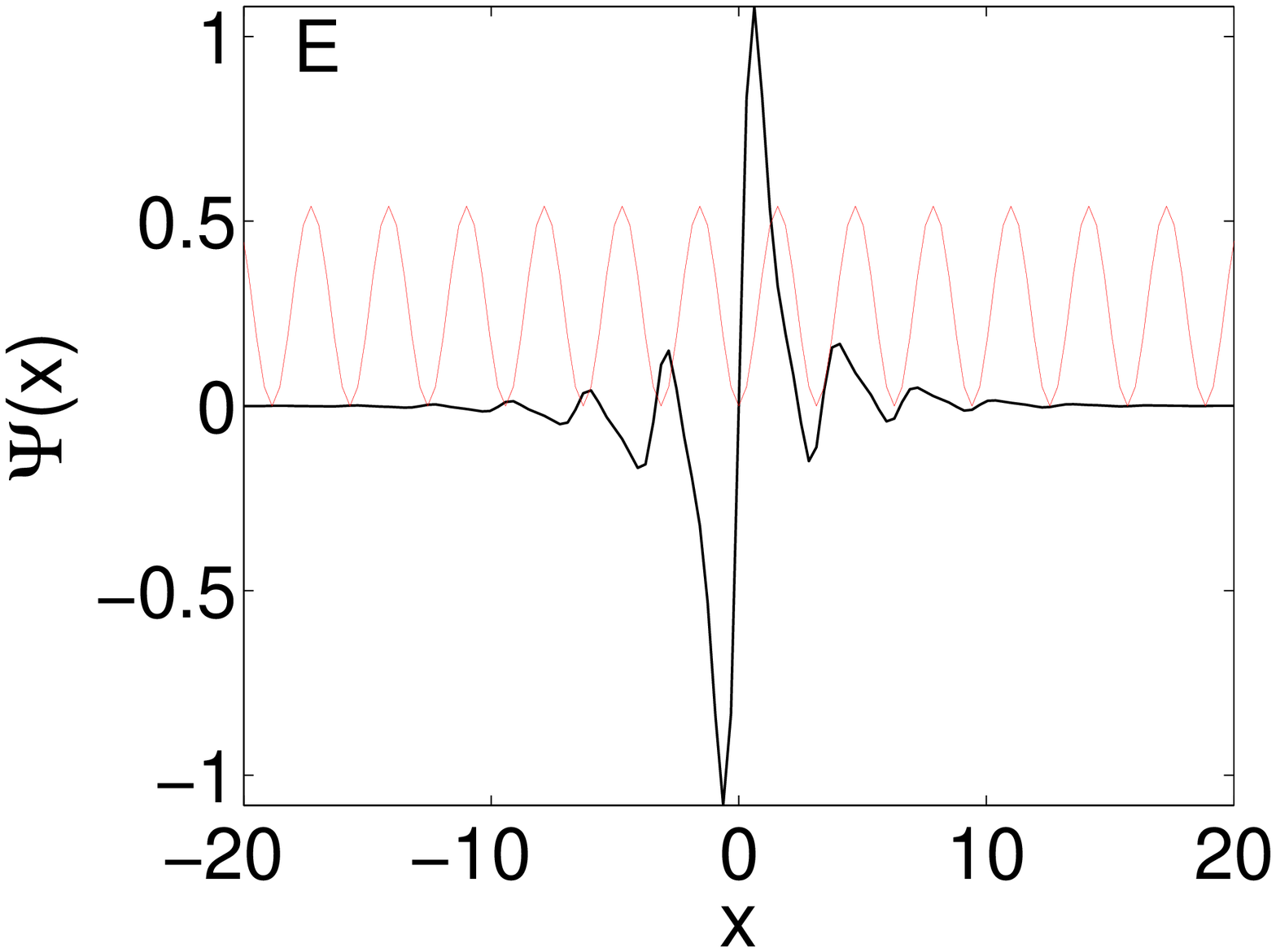} & \includegraphics[width=.2\textwidth]{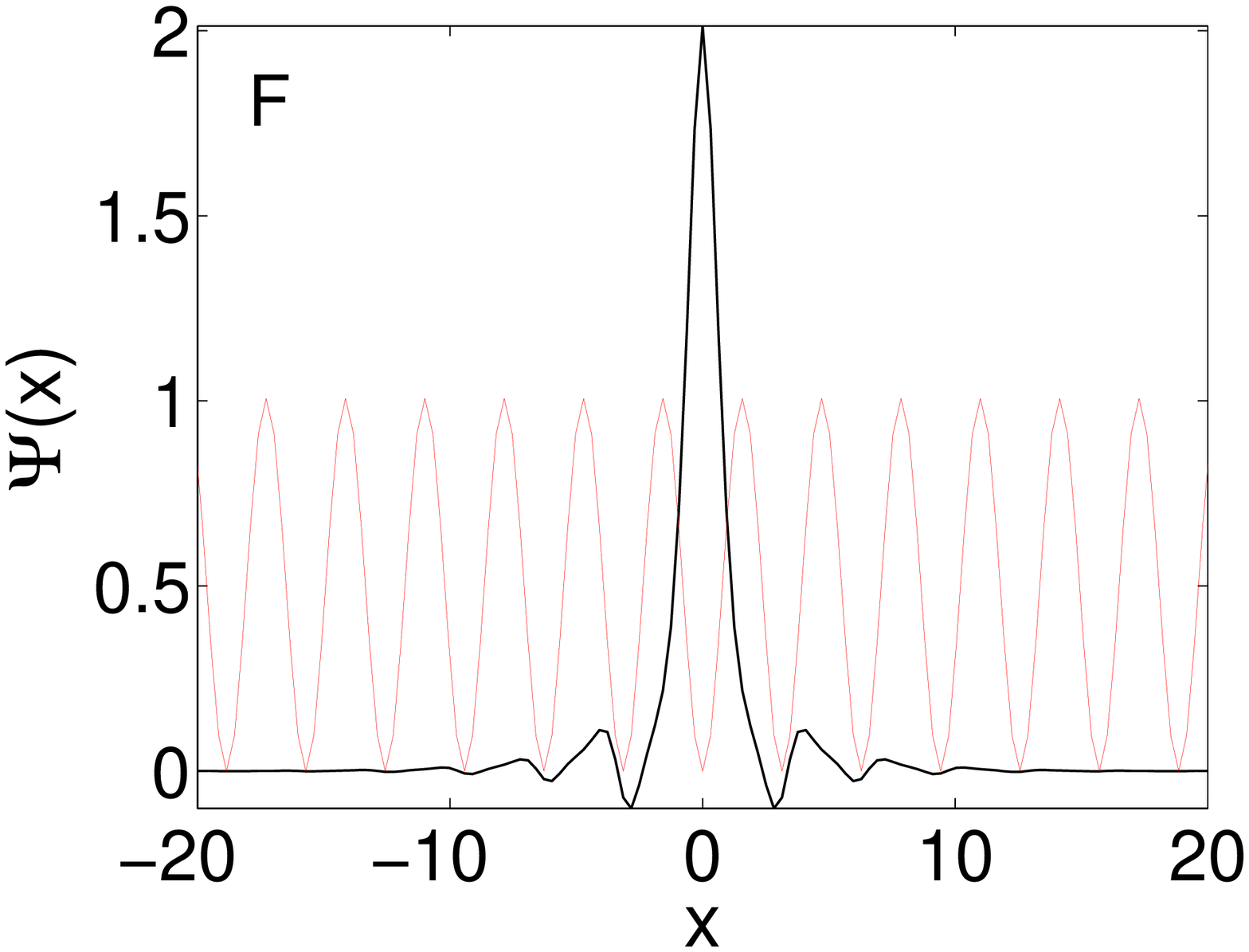} & \includegraphics[width=.2\textwidth]{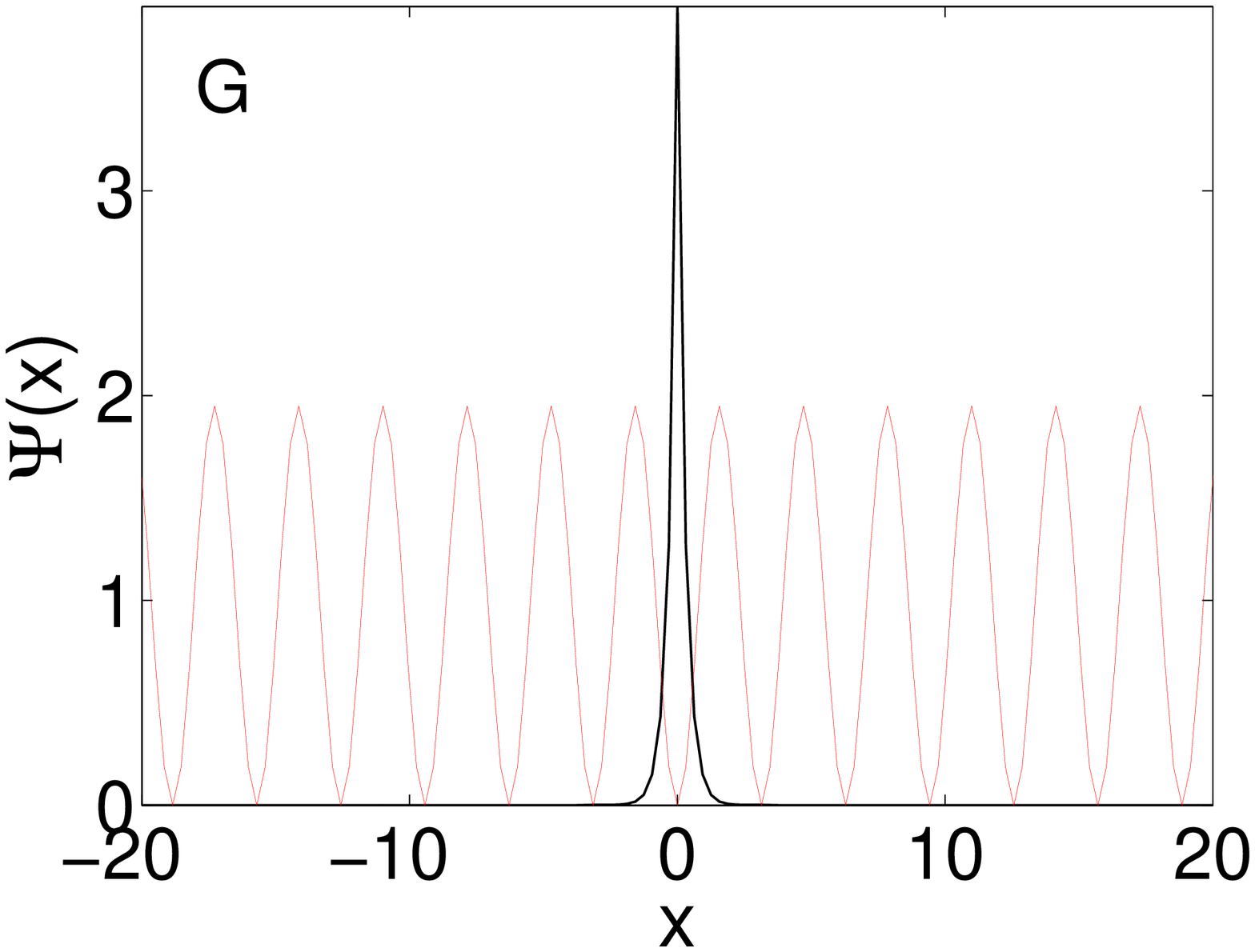} & 
\includegraphics[width=.2\textwidth]{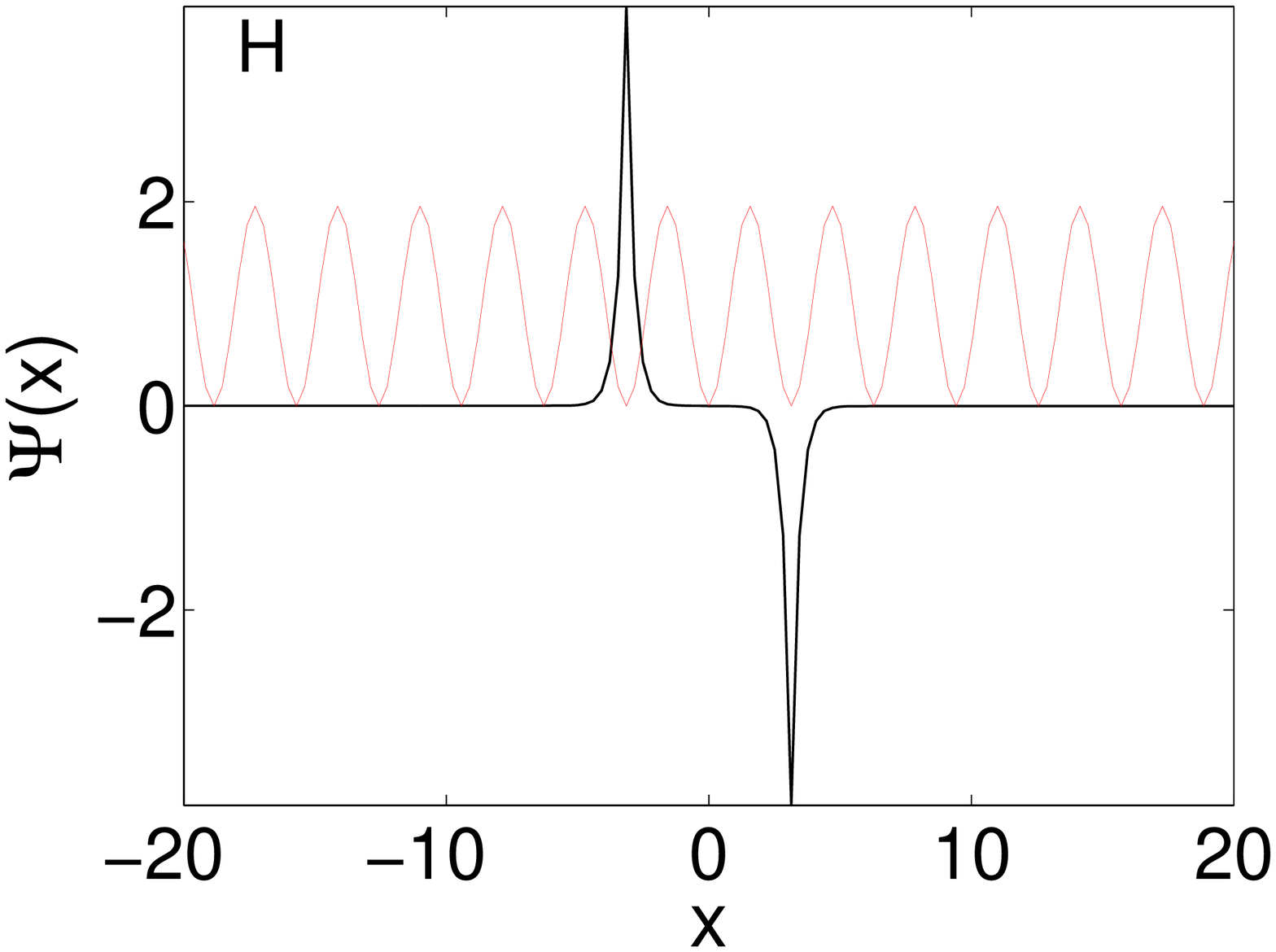} \\
& \includegraphics[width=.2\textwidth]{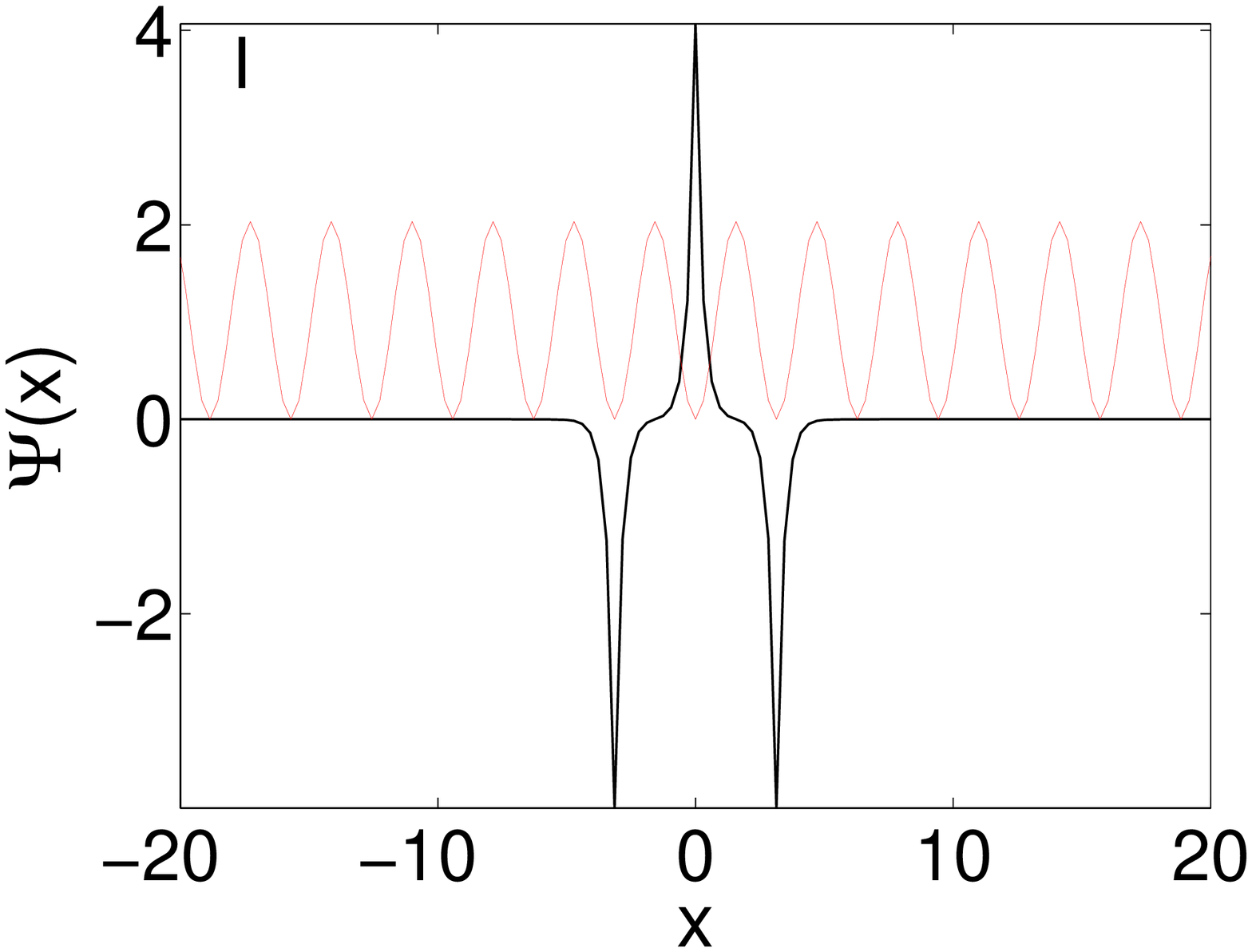} & \includegraphics[width=.2\textwidth]{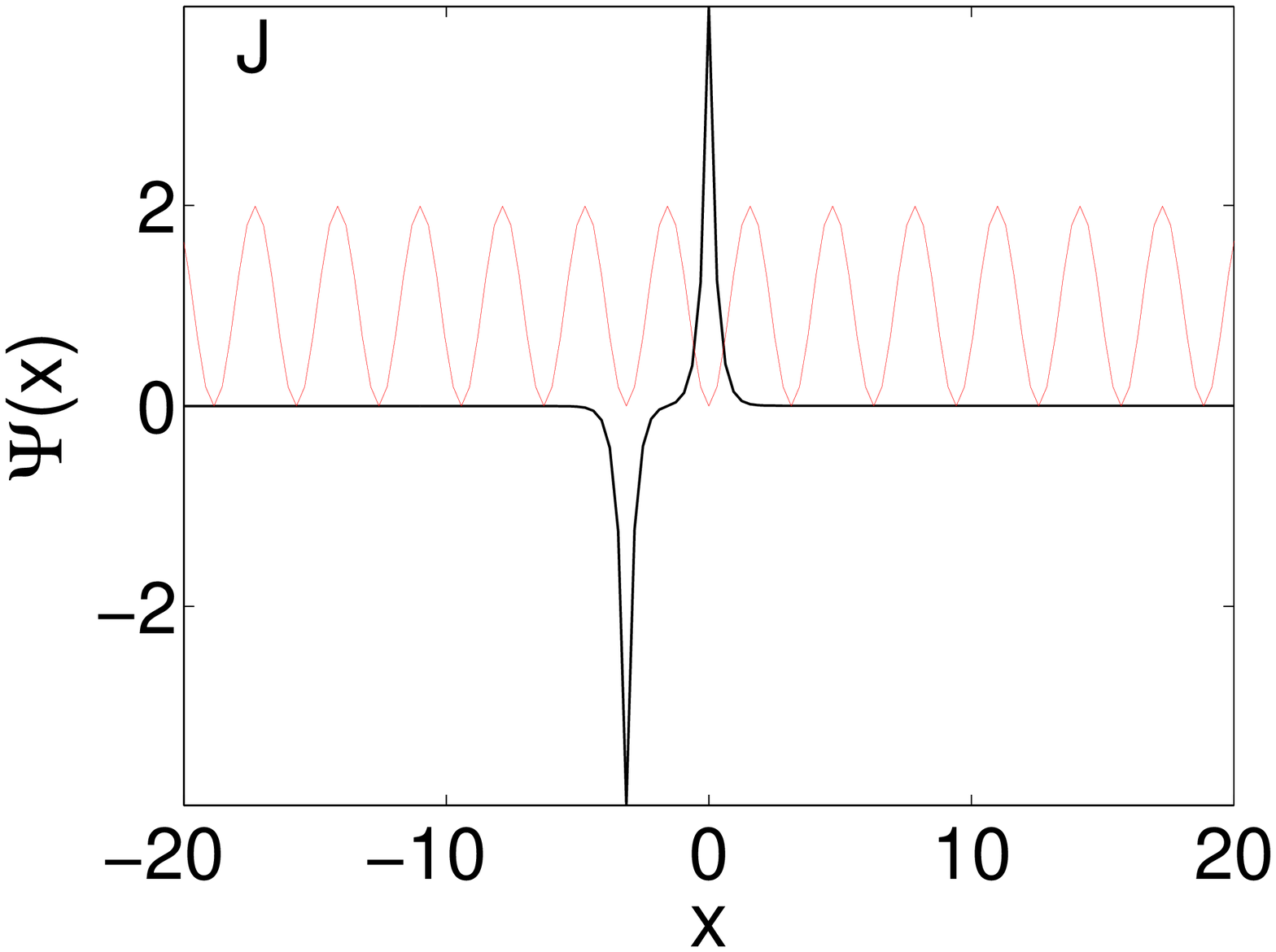} &  \\
&  &  &
\end{tabular}%
\end{center}
\caption{(Color online) Profiles of stable solitons labeled A-F in Fig.
\protect\ref{fig:normgapar}. Additionally, panels G-J display examples of
stable regular solitons found in the semi-infinite gap. The parameters are ($%
g_{c}=-1.5$, $\protect\mu =1$) and ($g_{c}=-1$, $\protect\mu =3~$and$\mathrm{%
~}5.5$) for the solitons in the semi-infinite gap, and first and second
finite bandgaps, respectively.}
\label{fig:profgapar}
\end{figure}

The existence regions for fundamental solitons (both regular ones in the
semi-infinite gap and GSs in the first two finite bandgaps) and SFSs in the (%
$\mu ,g_{c}$) plane are depicted together in Fig. \ref{fig:existar}; the
existence region for multi-humped bound states is not shown separately, as
it almost coincides with that of the fundamental solitons.\ In addition,
Fig. \ref{fig:normgapar} shows the norm as a function of $\mu $ at $g_{c}=-1$%
, for families of subfundamental and fundamental GSs, and for stable bound
states of fundamental GSs [in the semi-infinite gap, the norm of all types
of regular solitons very weakly depends on $\mu $, cf. Fig. \ref{fig:normar}%
(a)]. Typical examples of solitons of all these types are displayed in Fig. %
\ref{fig:profgapar}. Bound states in the second finite bandgap are not
shown, as they are completely unstable against oscillatory instabilities
(while they are stable in the semi-infinite and first finite gaps). In fact,
the same instability of bound states of GSs in the second finite bandgap
occurs in the local model.

\subsection{Repulsive local and attractive nonlocal interactions}

To adequately represent results in the model featuring the competition
between the local repulsion ($g_{c}>0$) and DD attraction ($g_{d}=-1$) in
the model with the OL, it was sufficient to use a coarser numerical mesh,
with $\Delta x=\pi /10$ (recall that $\Delta x=\pi /40$ was used above). The
nonlocal self-attraction allows the existence of solitons in the
semi-infinite gap. Similar to what was reported above for the same case in
the free-space model, the solitons become very broad at $g_{c}$ approaching
the critical value, $g_{c}\approx 3.6$. Other features of the regular
solitons found in the semi-infinite gap are also similar to those of their
free-space counterparts. The similarity holds also in the case of $g_{c}=0$,
i.e., in the model with the pure nonlocal attractive interactions.

\begin{figure}[tbp]
\begin{center}
\includegraphics[width=.5\textwidth]{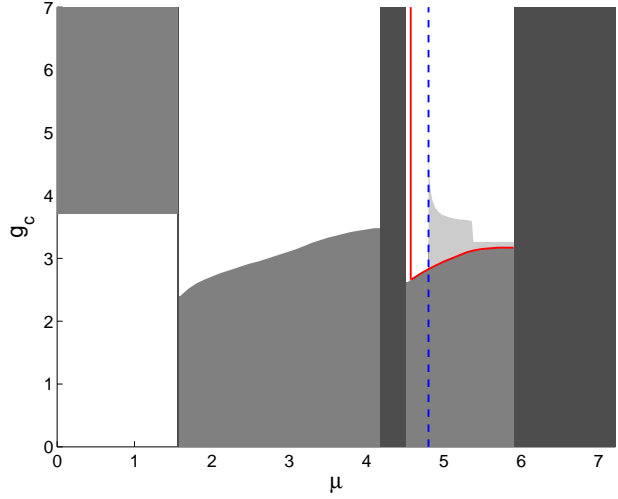}
\end{center}
\caption{(Color online) The existence range for solitons in the
semi-infinite and two lowest finite gaps, in the model with the optical
lattice and competing local repulsive and nonlocal attractive interactions.
The notation is the same as in Fig. \protect\ref{fig:existar}.}
\label{fig:existra}
\end{figure}

\begin{figure}[tbp]
\begin{center}
\begin{tabular}{cc}
\includegraphics[width=.5\textwidth]{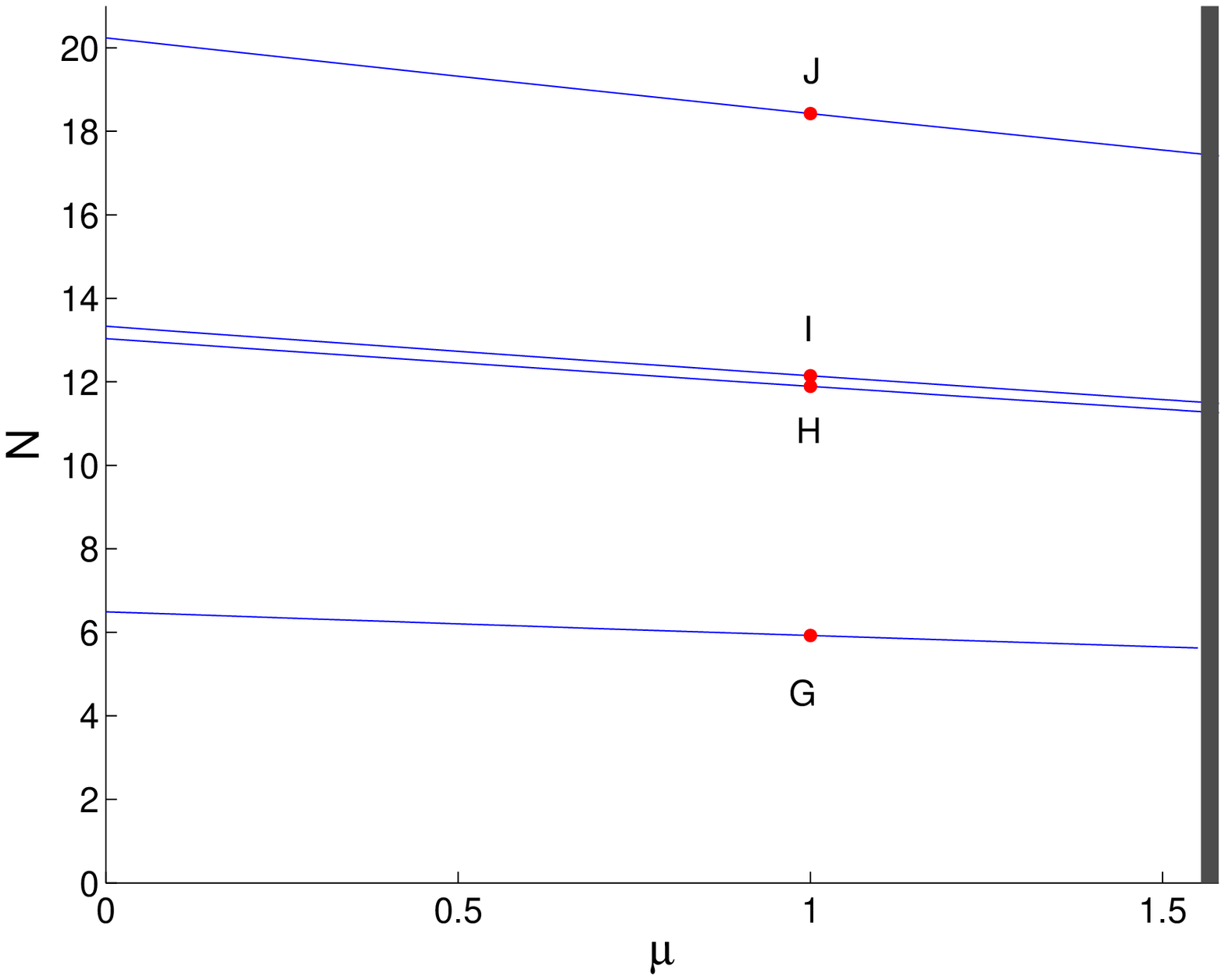} & \includegraphics[width=.5\textwidth]{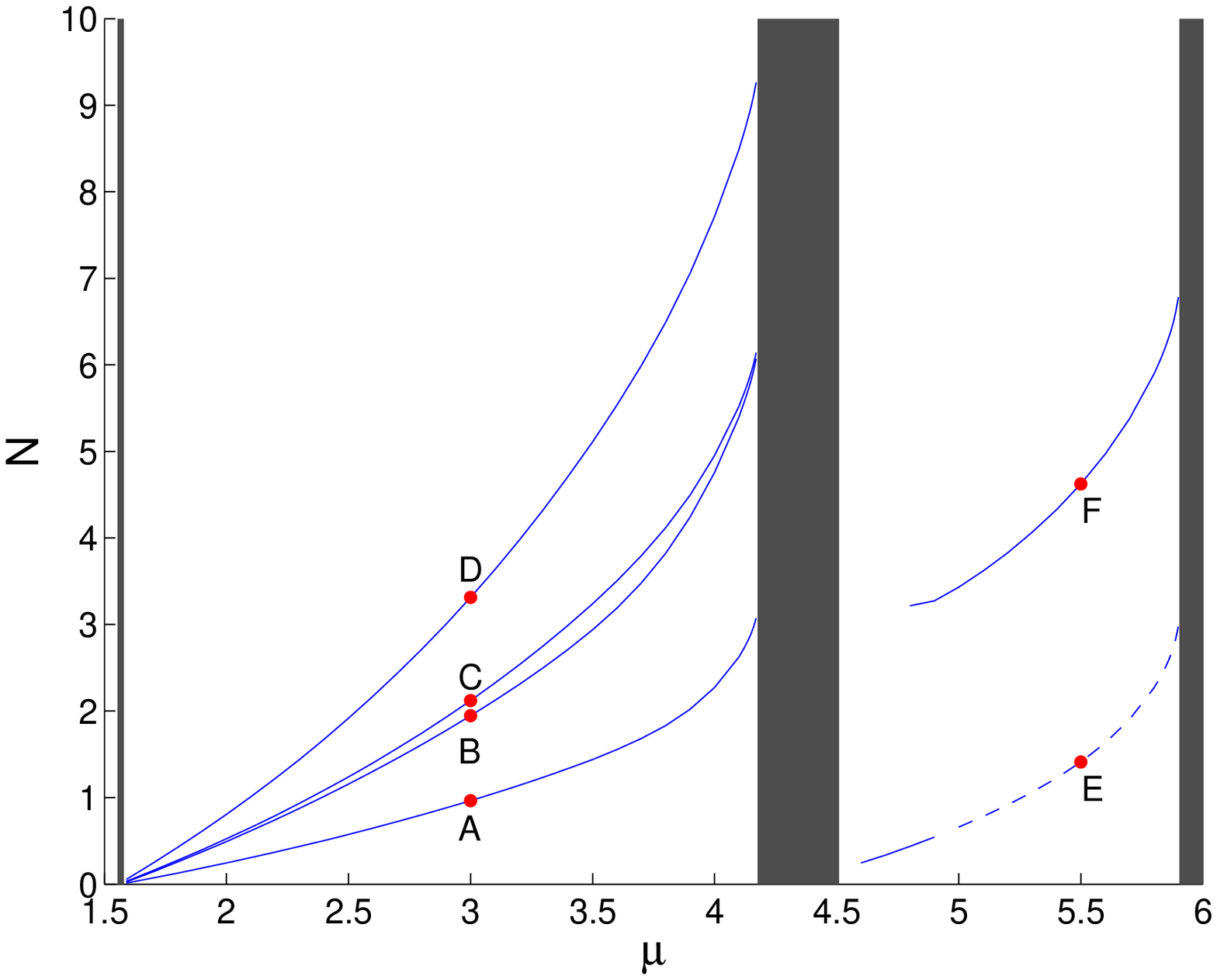} \\
&
\end{tabular}%
\end{center}
\caption{(Color online) $N(\protect\mu )$ curves for regular (left panel)
and gap (right panel) solitons in the model with the optical lattice, local
repulsion, and nonlocal attraction. The strength of the local repulsion is
fixed to $g_{c}=1$ (a) and $g_{c}=5$ (b).}
\label{fig:normgapra}
\end{figure}

\begin{figure}[tbp]
\begin{center}
\begin{tabular}{cccc}
\includegraphics[width=.2\textwidth]{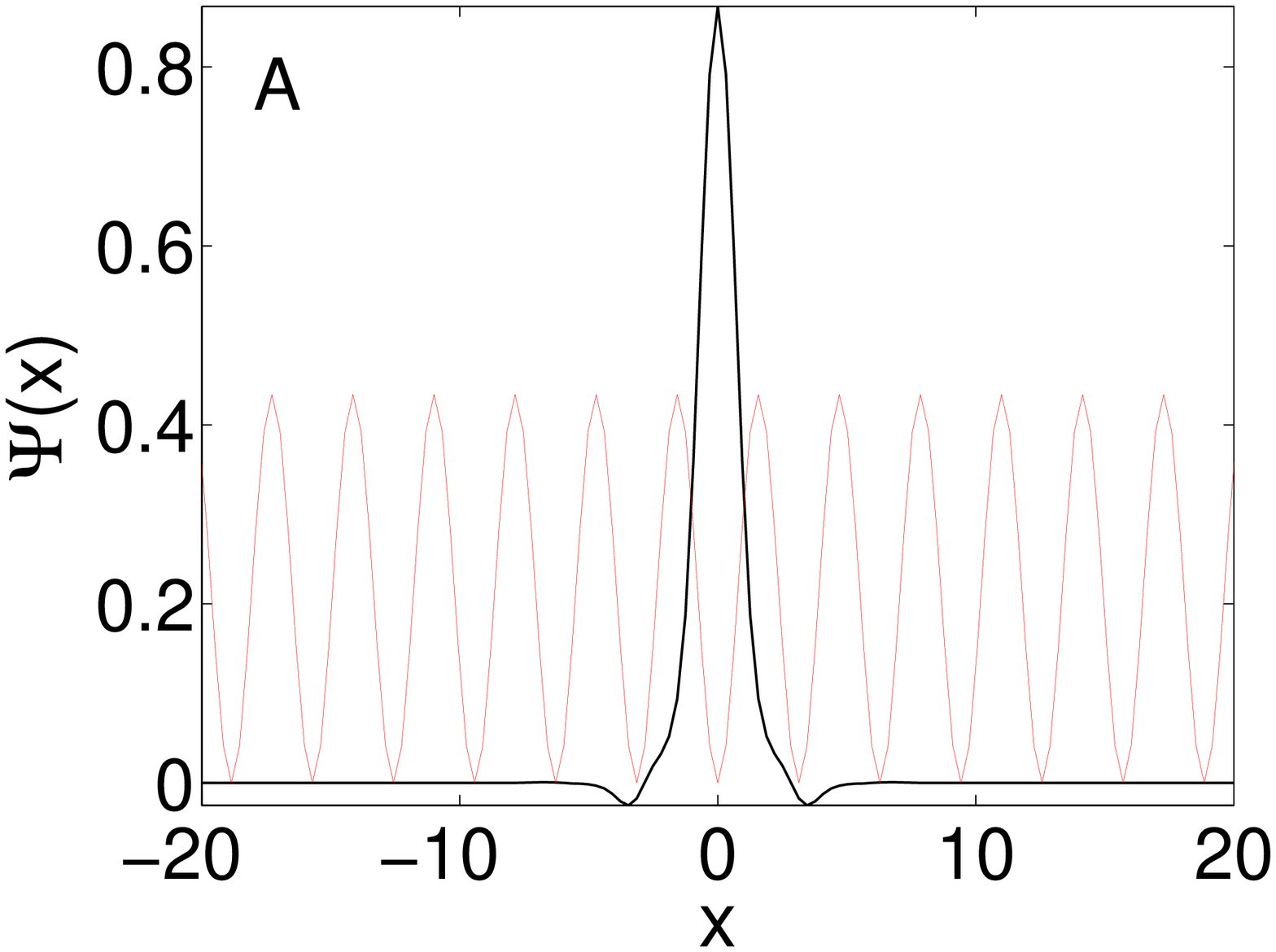} & \includegraphics[width=.2\textwidth]{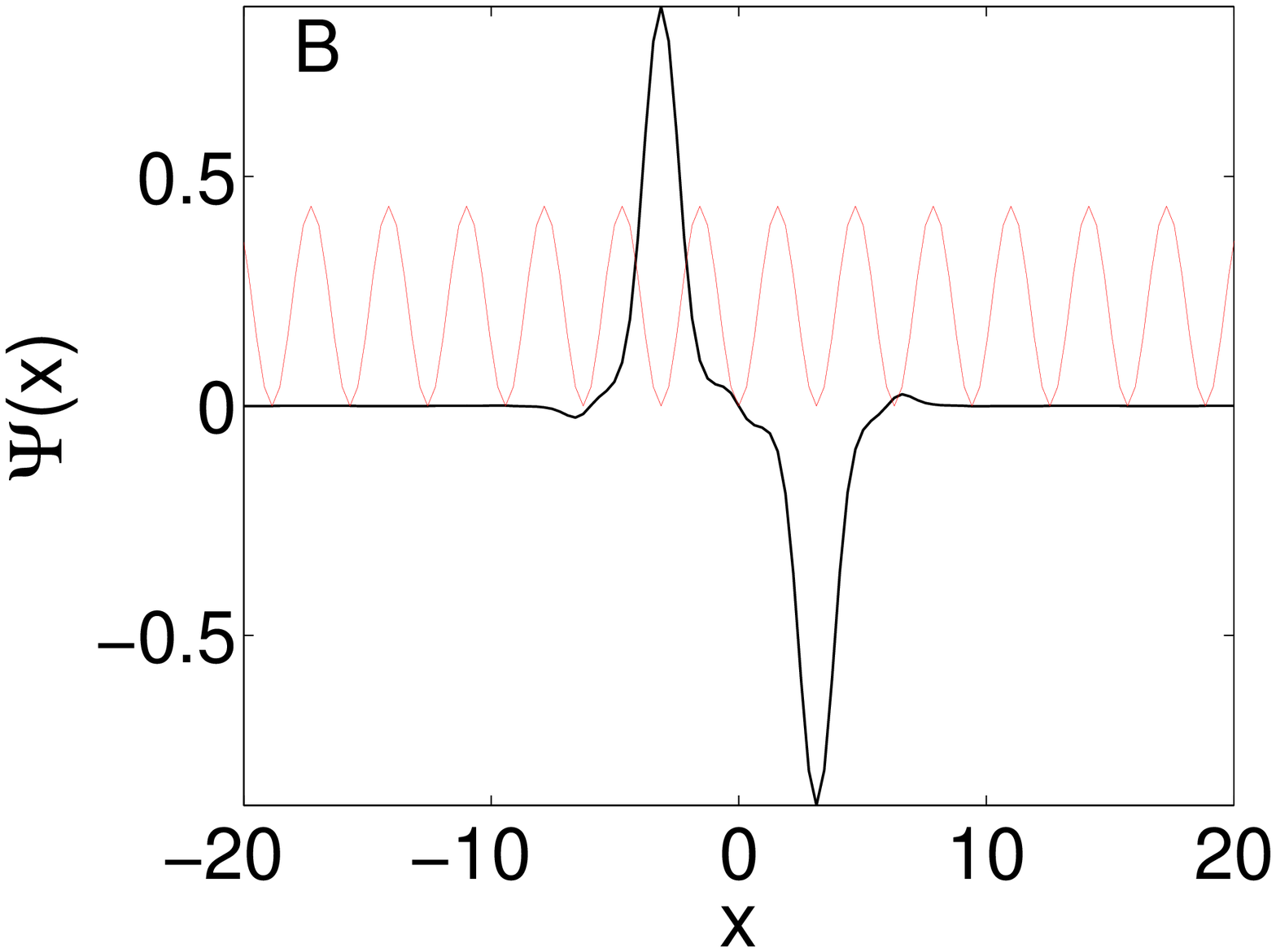} & \includegraphics[width=.2\textwidth]{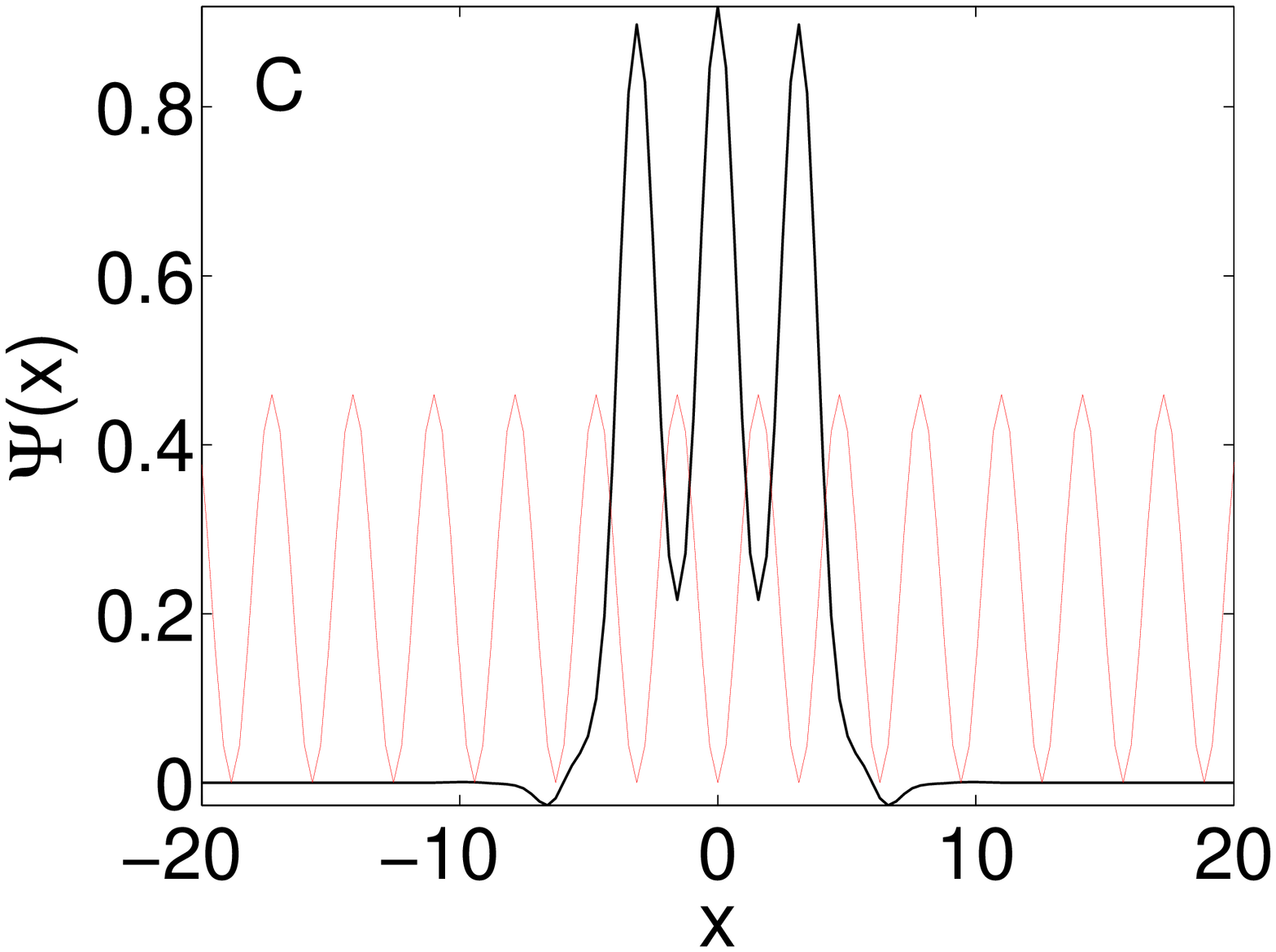} & %
\includegraphics[width=.2\textwidth]{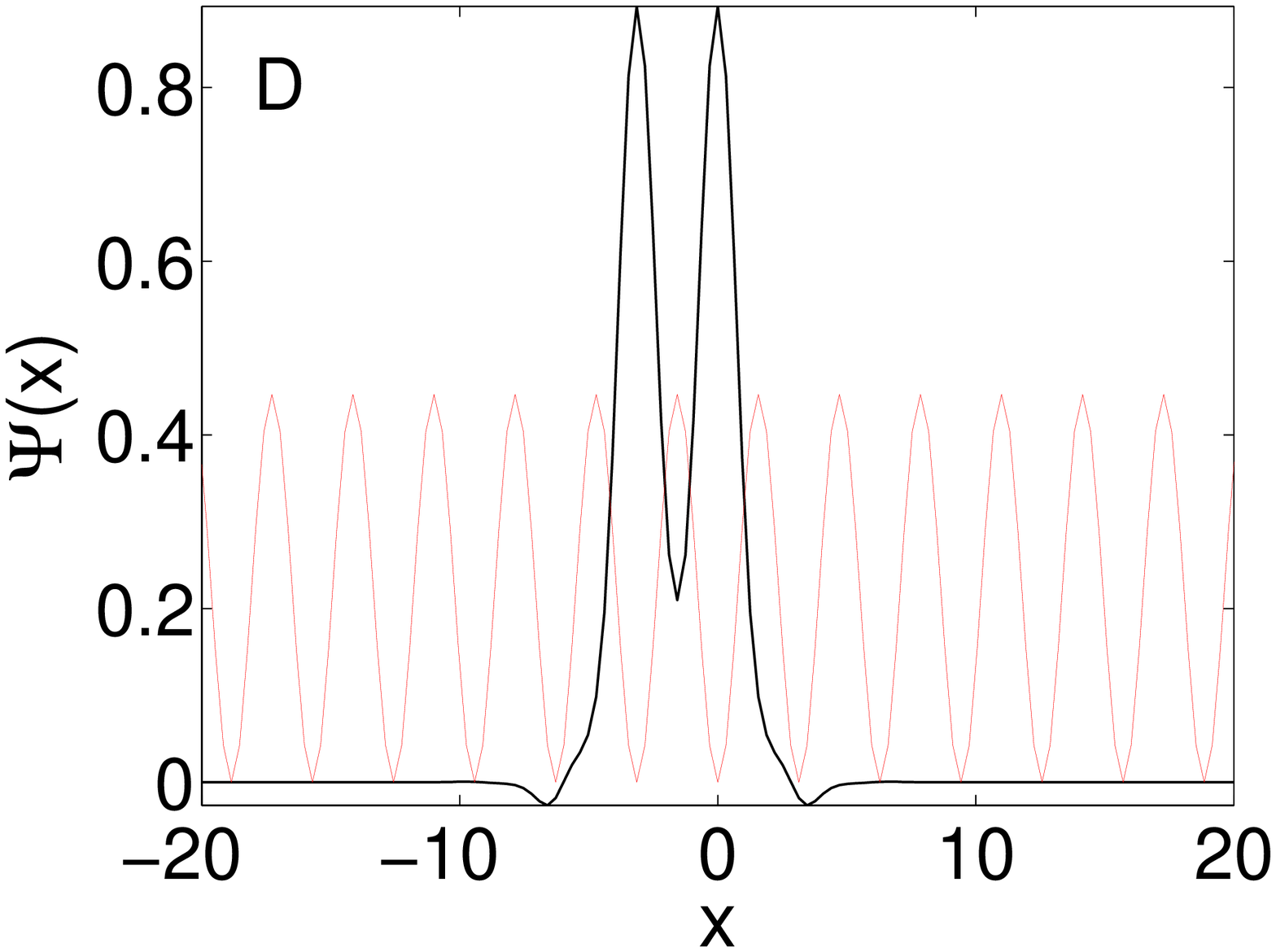} \\
\includegraphics[width=.2\textwidth]{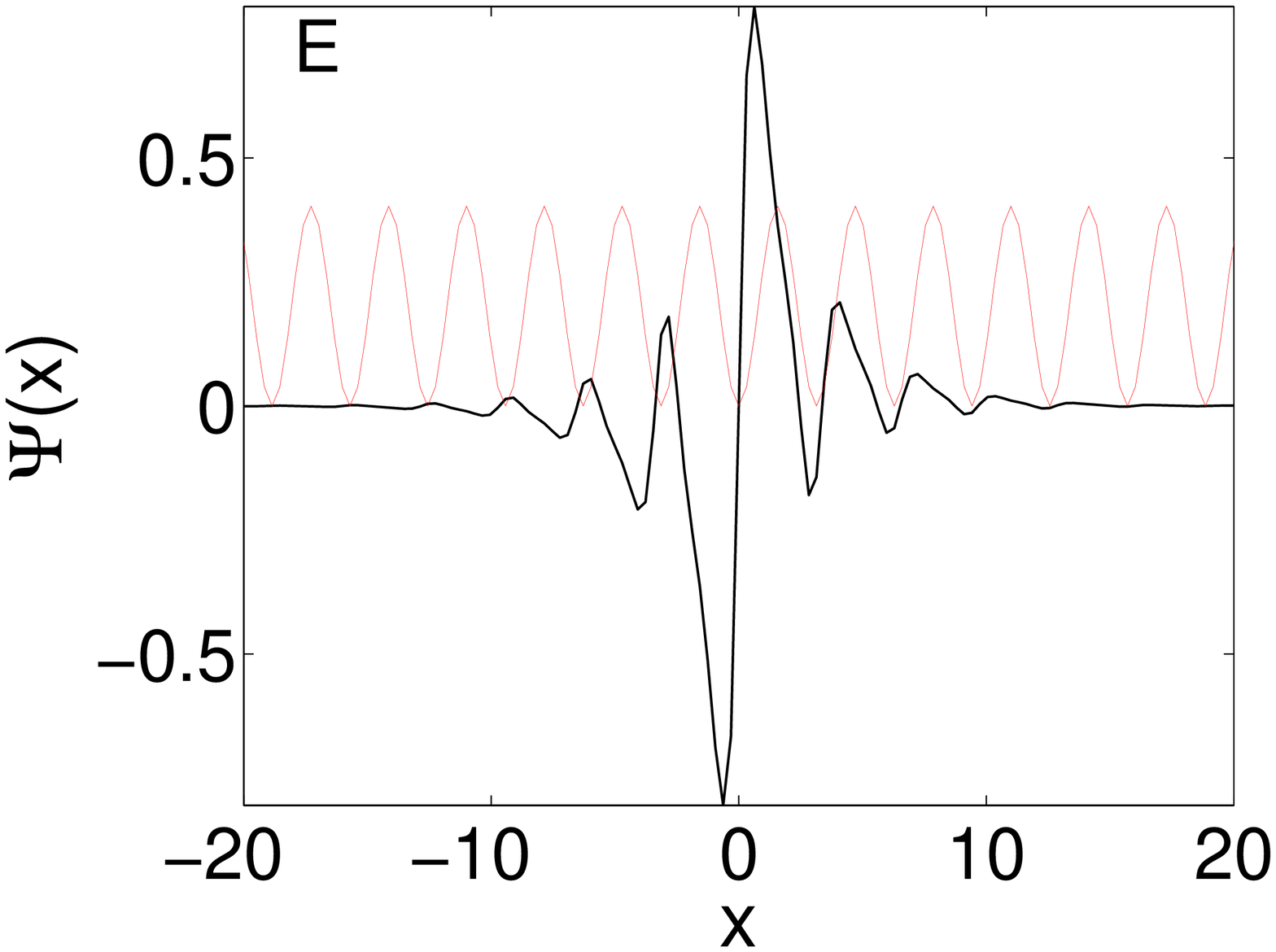} & \includegraphics[width=.2\textwidth]{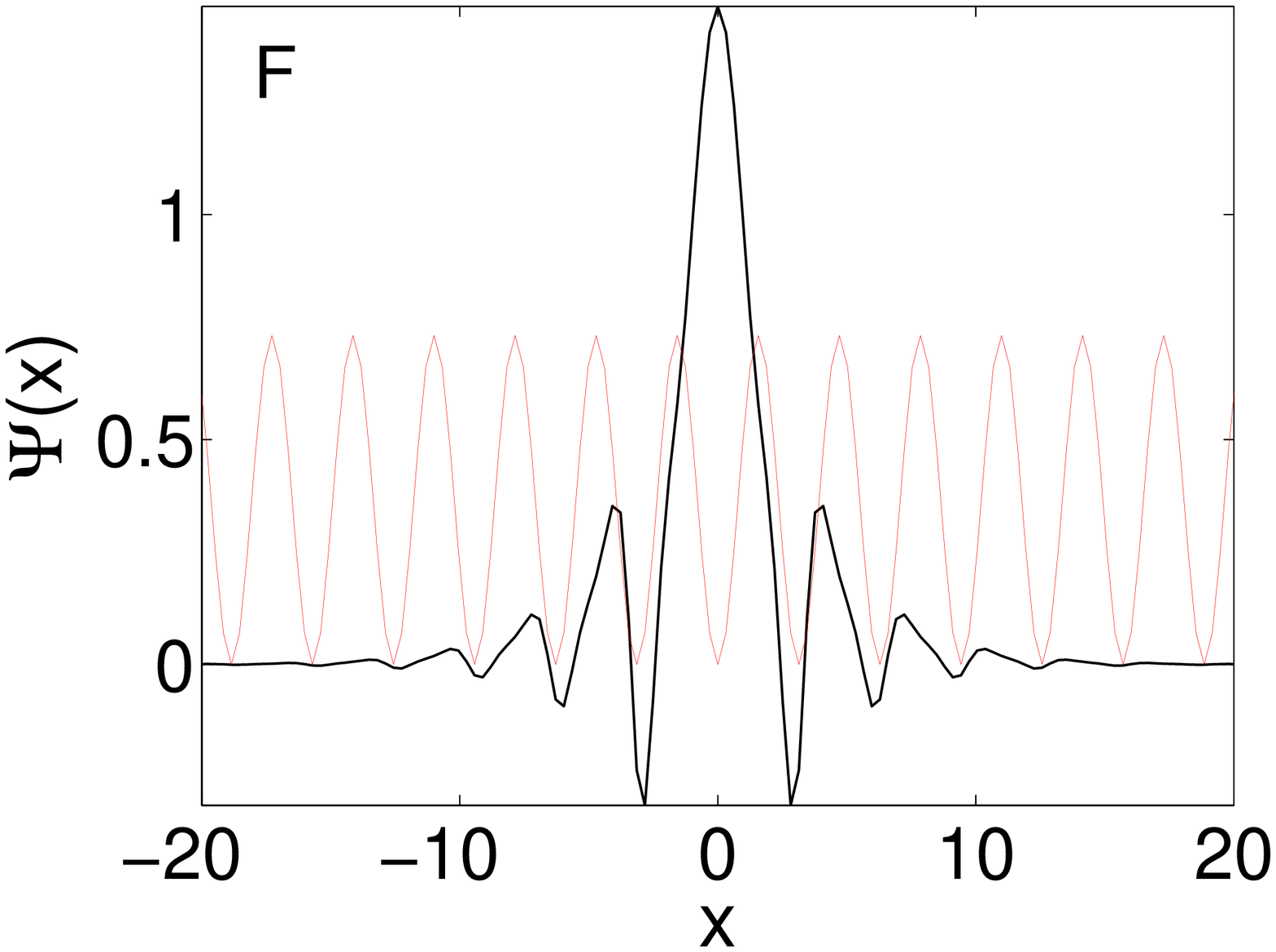} & \includegraphics[width=.2\textwidth]{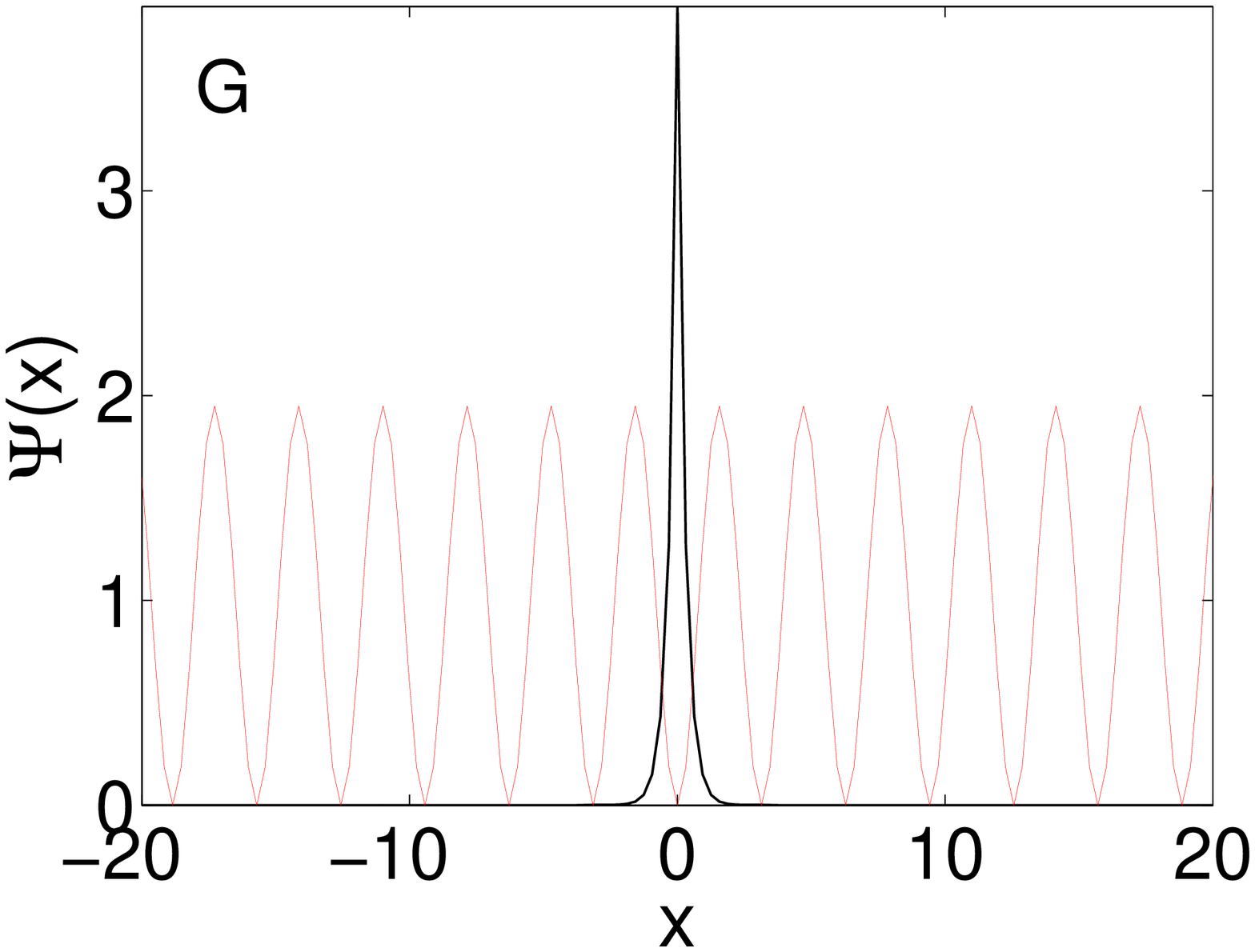} & %
\includegraphics[width=.2\textwidth]{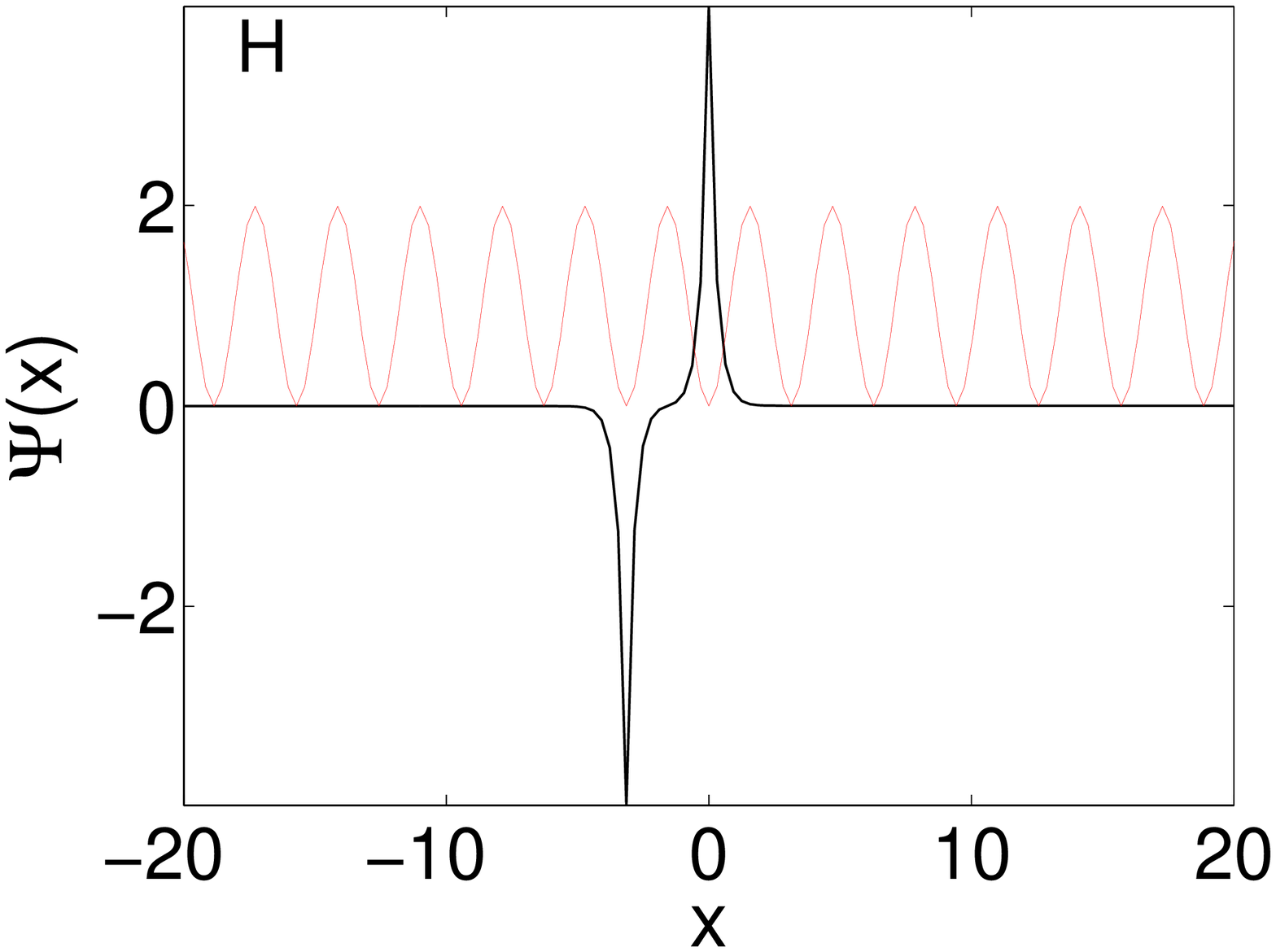} \\
& \includegraphics[width=.2\textwidth]{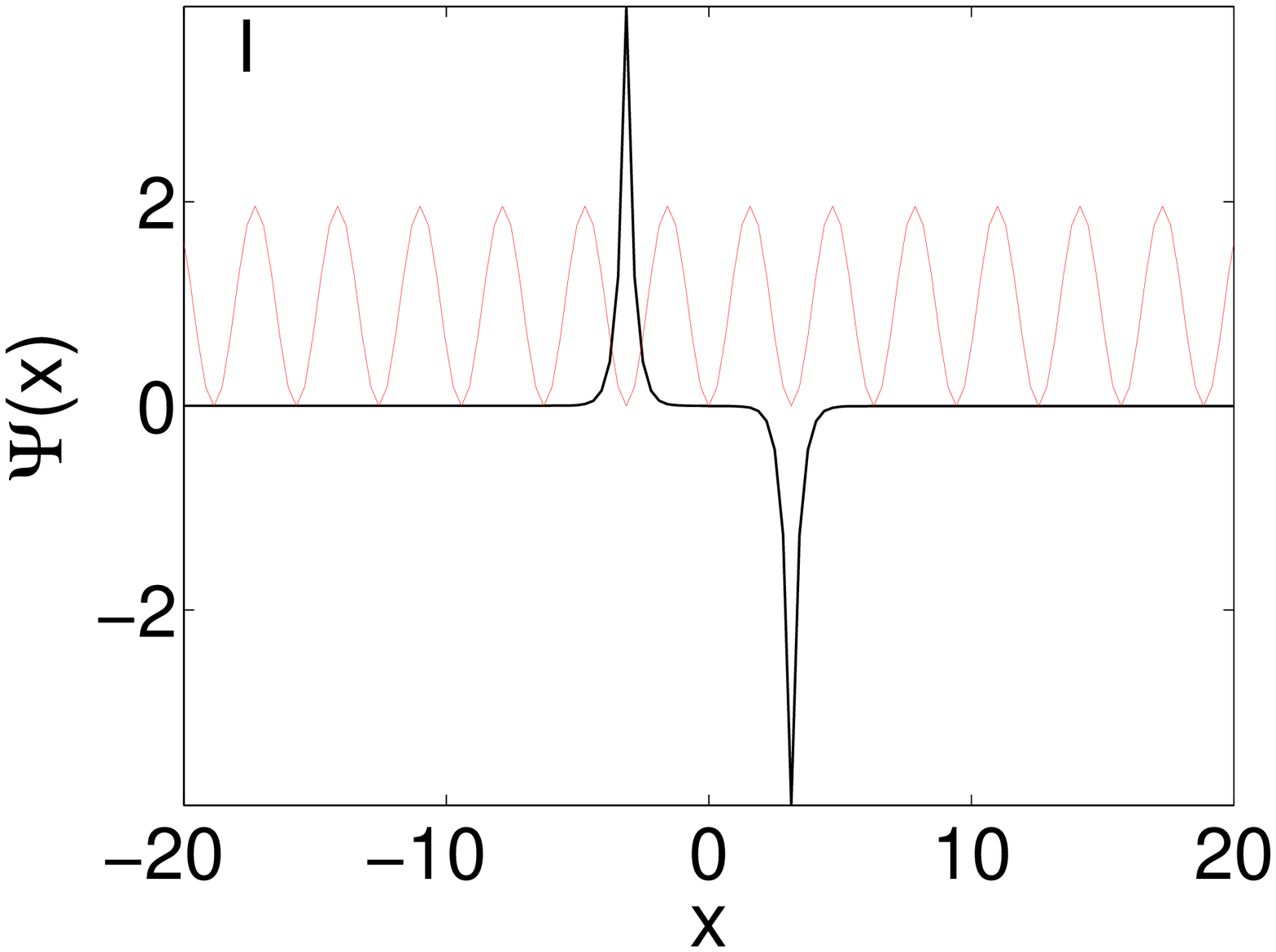} & %
\includegraphics[width=.2\textwidth]{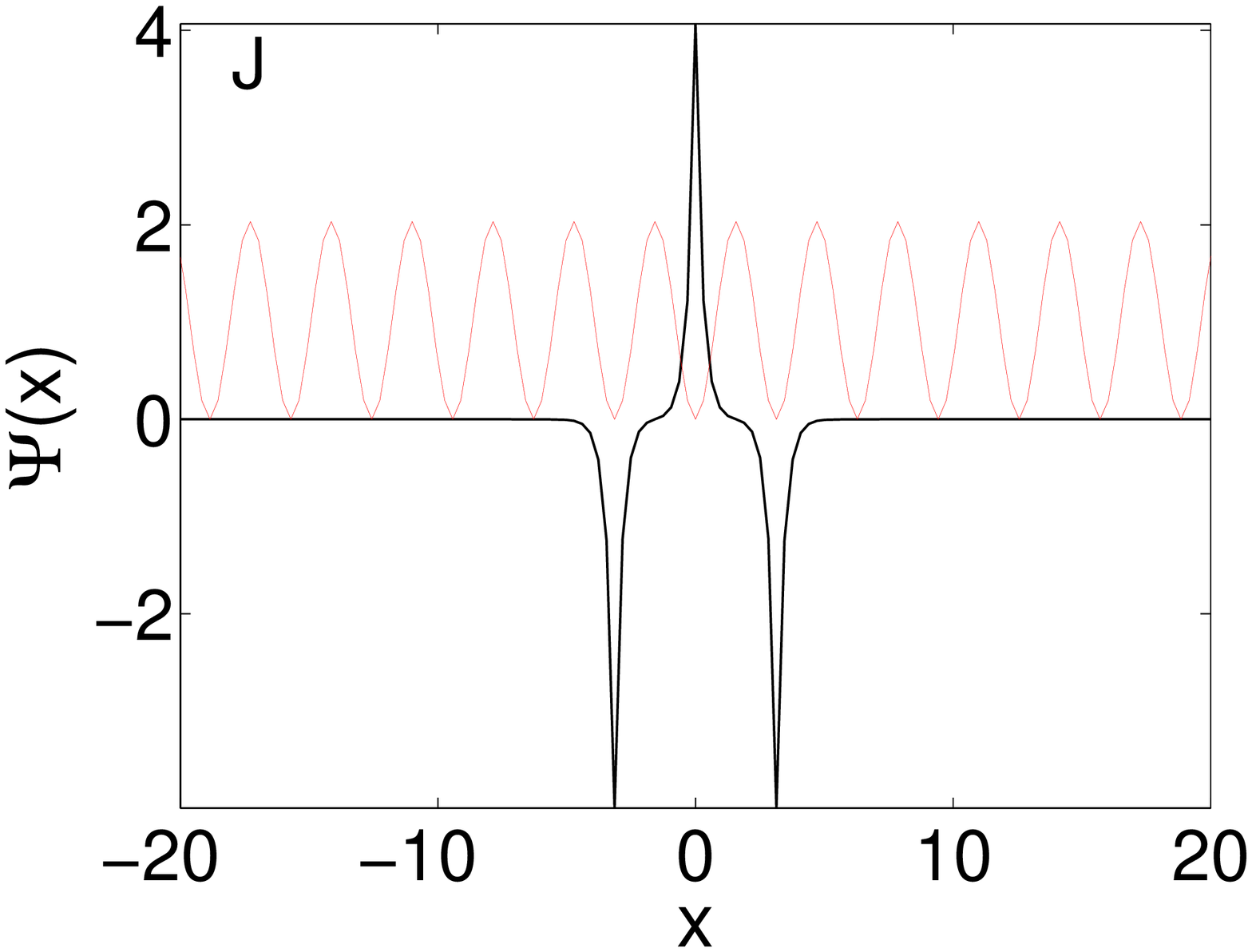} &  \\
&  &  &
\end{tabular}%
\end{center}
\caption{(Color online) Soliton profiles at points A-J in Fig. \protect\ref%
{fig:normgapra}. The parameters are ($g_{c}=1$, $\protect\mu =1$) and ($%
g_{c}=5$, $\protect\mu =3~$and$\mathrm{~}5.5$) for the solitons in the
semi-infinite gap, and first and second finite bandgaps, respectively.}
\label{fig:profgapra}
\end{figure}

The existence range for stable fundamental solitons (both the regular ones
and GSs) and SFSs in the present version of the model is depicted in Fig. %
\ref{fig:existra}. Further, Fig. \ref{fig:normgapra} shows the respective $%
N(\mu )$ dependences, including those for bound-state solutions. This figure
shows $N(\mu )$ lines in the semi-infinite gap too, as, on the contrary to
the situation for the model with $g_{c}<0$ and $g_{d}=1$, these lines do not
degenerate into $N=\mathrm{const}$. In fact, the opposite signs of $dN/d\mu $
for regular solitons and GSs is a typical feature, observed in local models
as well. Typical profiles of various soliton species belonging to the
semi-infinite and finite gaps are displayed in Fig. \ref{fig:profgapra}.

\section{Collisions between moving solitons}

\begin{figure}[tbp]
\begin{center}
\begin{tabular}{cc}
\includegraphics[width=.5\textwidth]{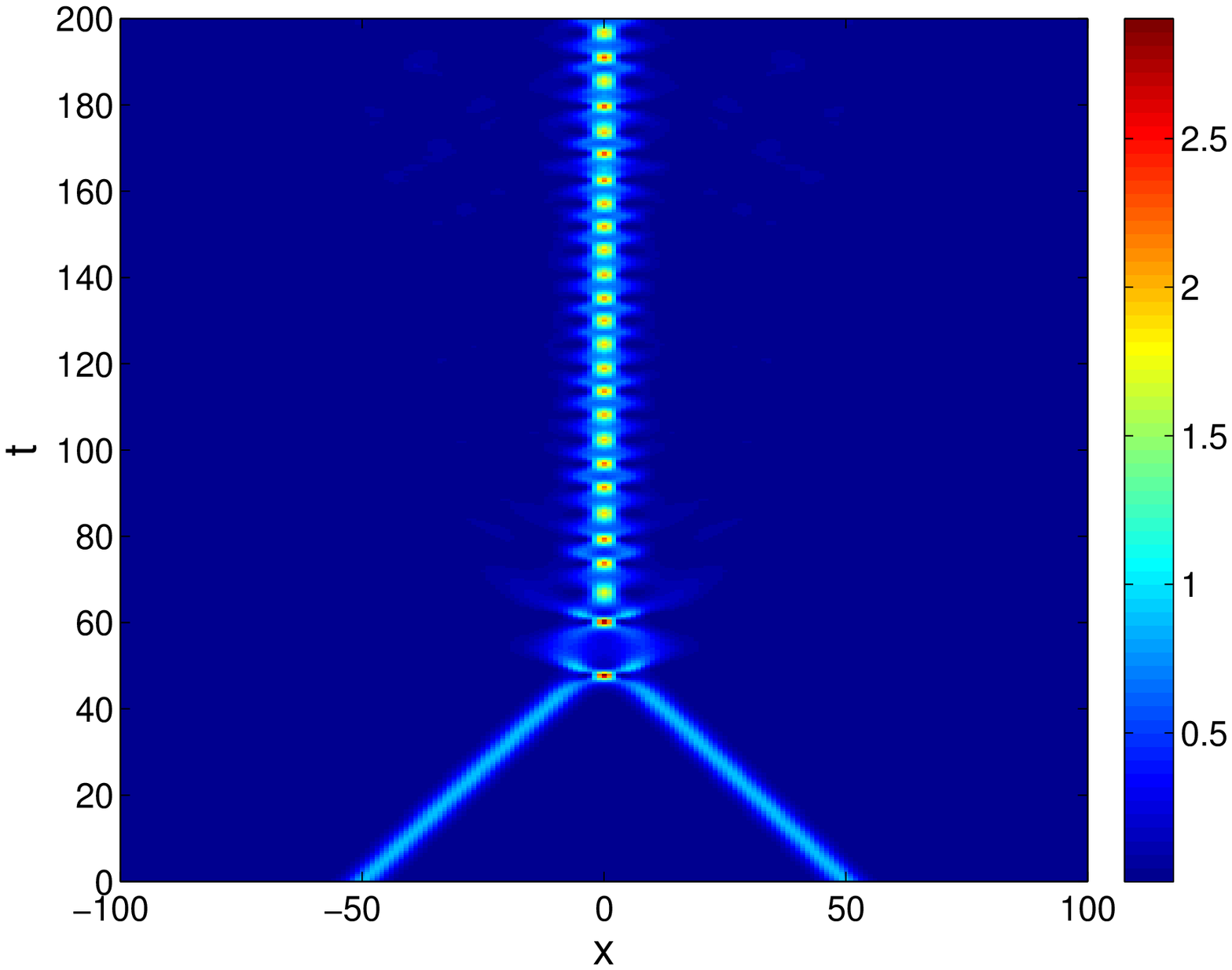} & \includegraphics[width=.5\textwidth]{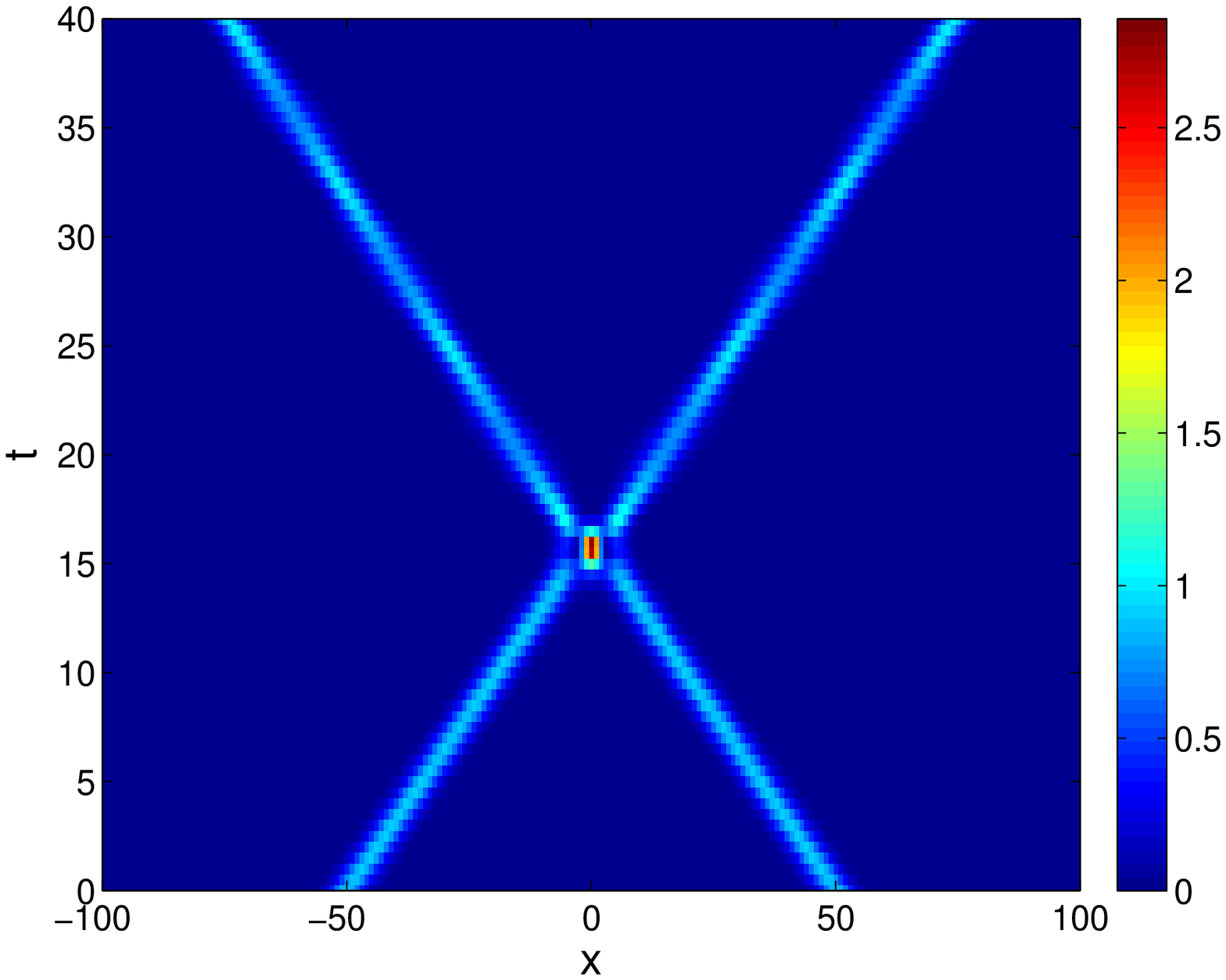} \\
(a) & (b)%
\end{tabular}%
\end{center}
\caption{(Color online) Soliton-soliton collisions in the free space ($%
\protect\epsilon =0$) for $g_{c}=1$, $g_{d}=-1$, and $\protect\mu =-1$. The
initial velocities are $c=\pm 0.2942$ (a) and $c=\pm 0.9651$ (b).}
\label{fig:collra}
\end{figure}

We have also studied collisions between solitons moving in the free space.
In the model with the repulsive local and attractive DD interactions, a
usual collision scenario is observed: at small velocities, solitons merge
into a bound state, while, at high velocities, they pass through each other,
as shown in Fig. \ref{fig:collra}.

\begin{figure}[tbp]
\begin{center}
$%
\begin{array}{ccc}
\includegraphics[width=.32\textwidth]{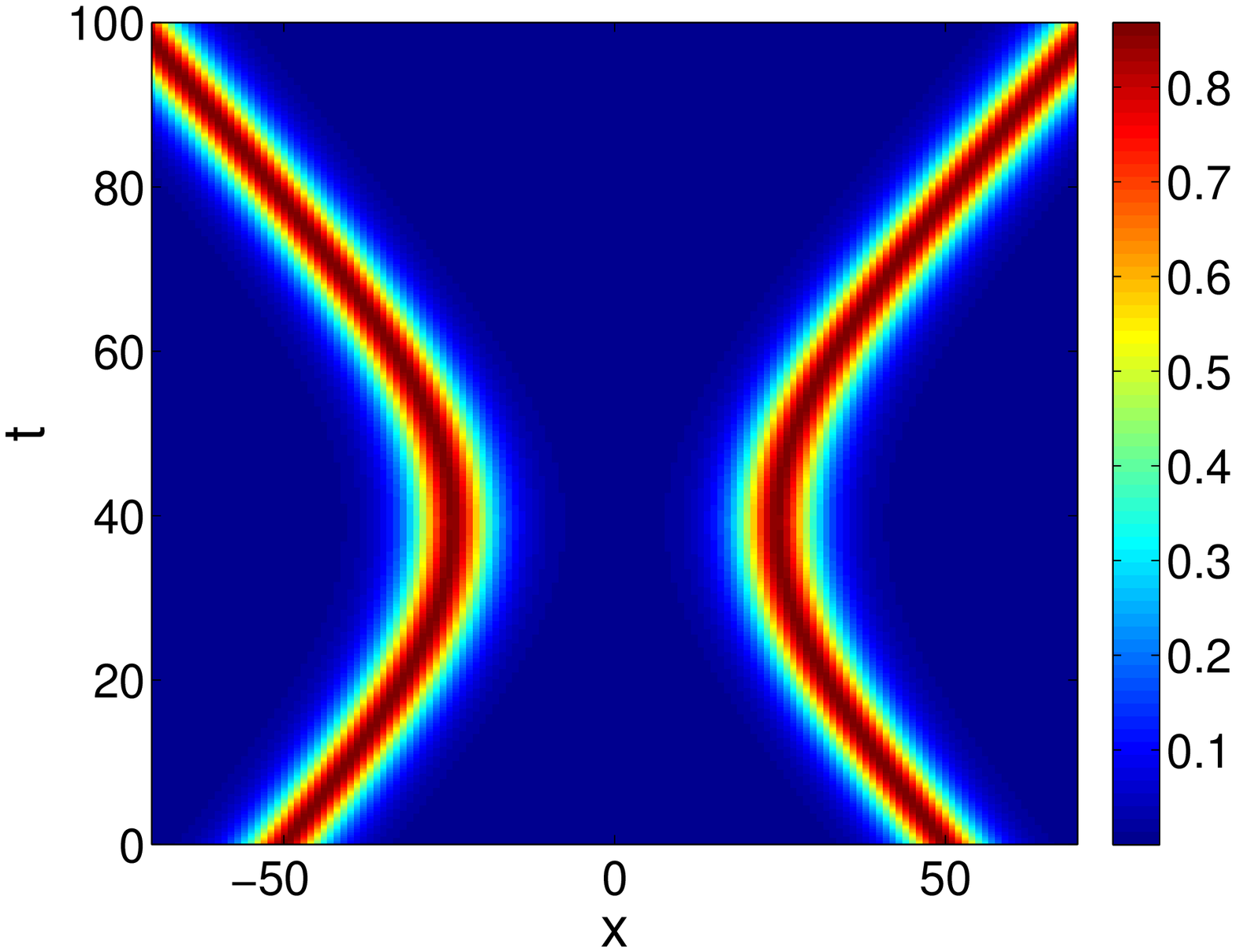} & %
\includegraphics[width=.32\textwidth]{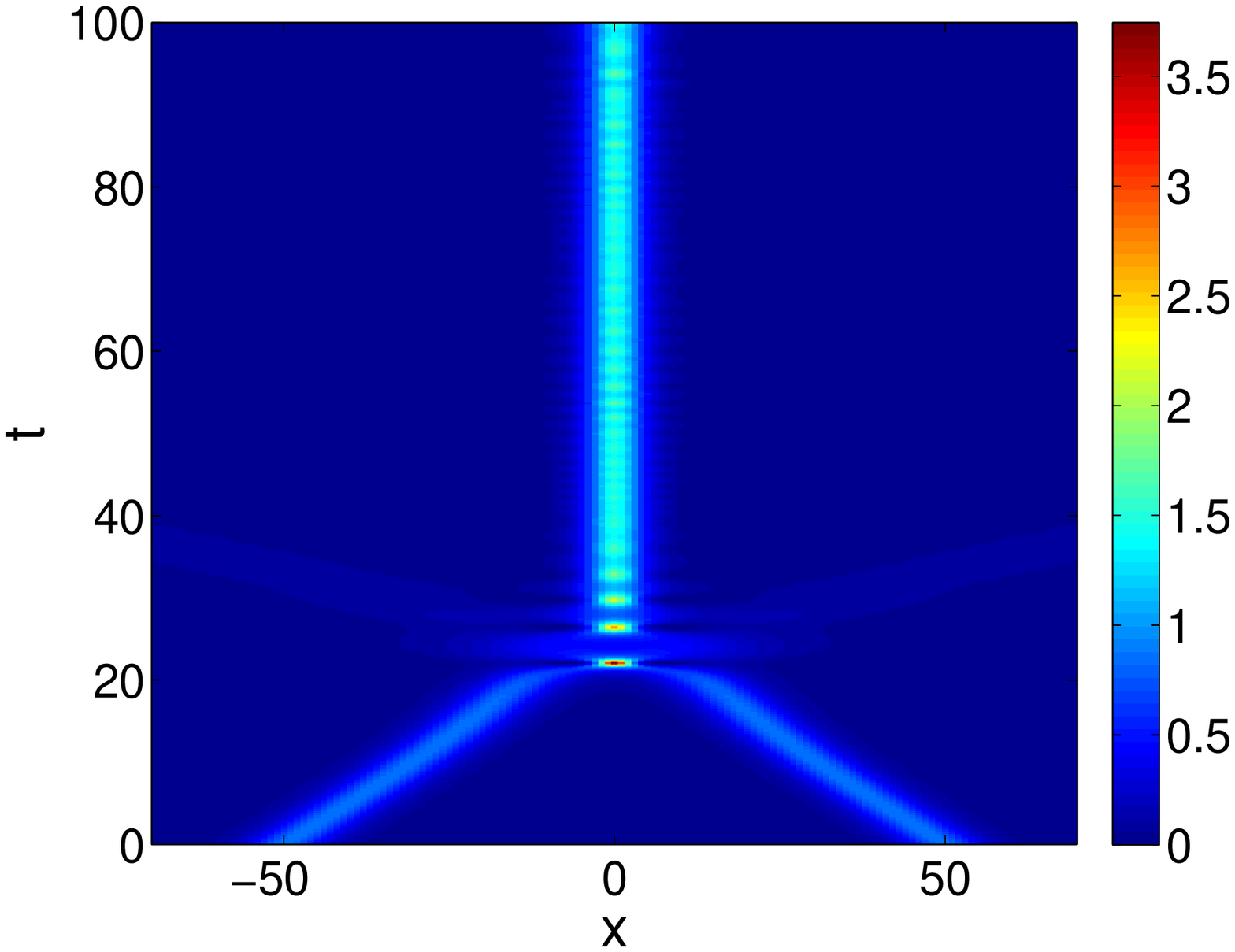} & %
\includegraphics[width=.32\textwidth]{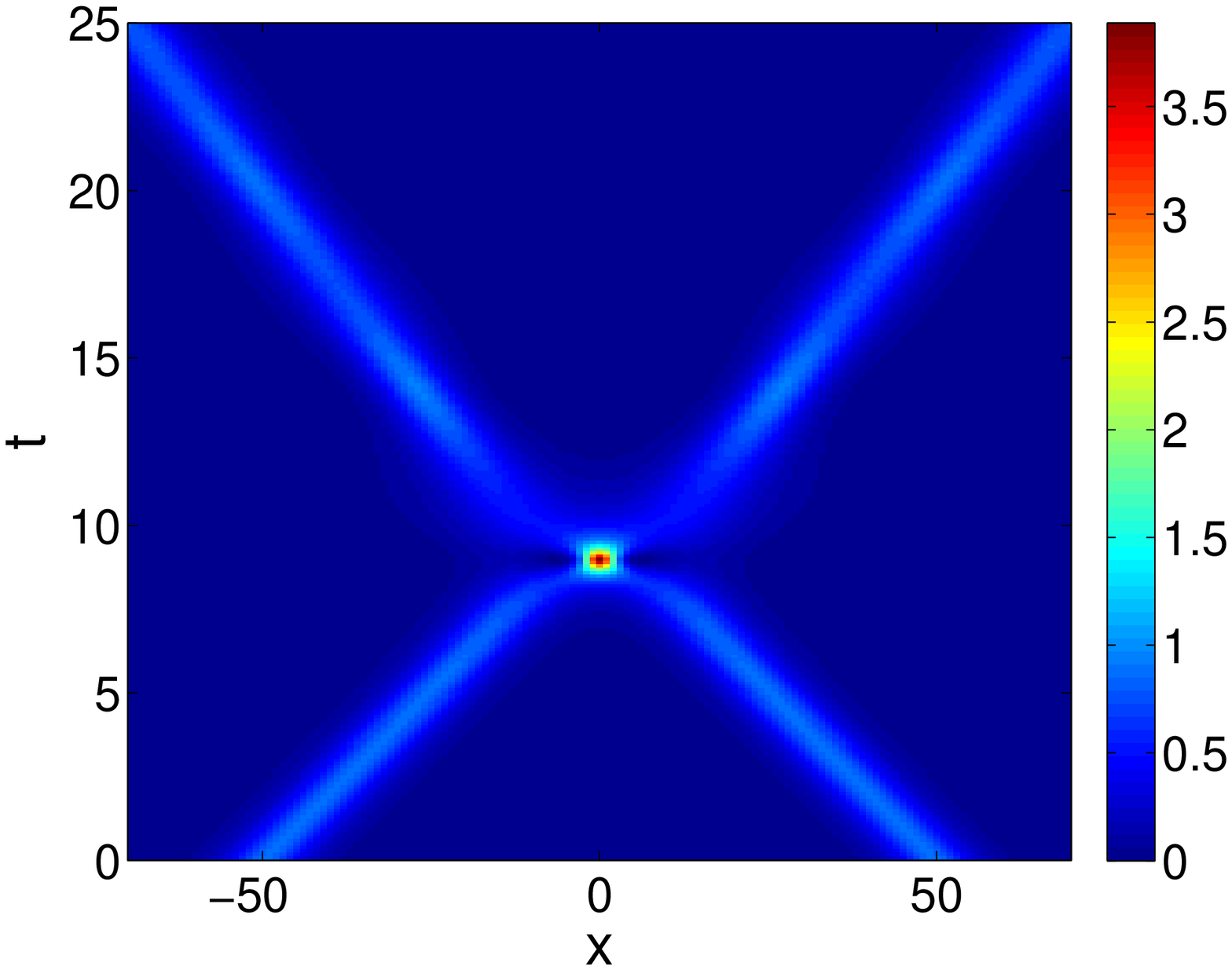} \\
(\mathrm{a}) & (\mathrm{b}) & (\mathrm{c})%
\end{array}%
$%
\end{center}
\caption{(Color online) Soliton collisions at $g_{c}=-5$, $g_{d}=1$, $%
\protect\mu =-1$ and $\protect\epsilon =0$. Initial velocities are $c=\pm
0.0990$ (a), $\pm 0.1984$ (b), and $\pm 0.4969$ (c). These plots were
generated using the mesh with $\Delta x=0.1$.}
\label{fig:collar}
\end{figure}

The collision scenario is different in the opposite case, with the local
attraction and nonlocal repulsion. As shown in Fig. \ref{fig:collar}, at
small velocities the solitons bounce from each other. This feature is easily
explained by the fact that the long-range interaction between the solitons
is repulsive. The rebound is changed by the merger at intermediate values of
the velocities. Finally, fast solitons pass through each other.

\begin{figure}[tbp]
\begin{center}
\begin{tabular}{cc}
\includegraphics[width=.5\textwidth]{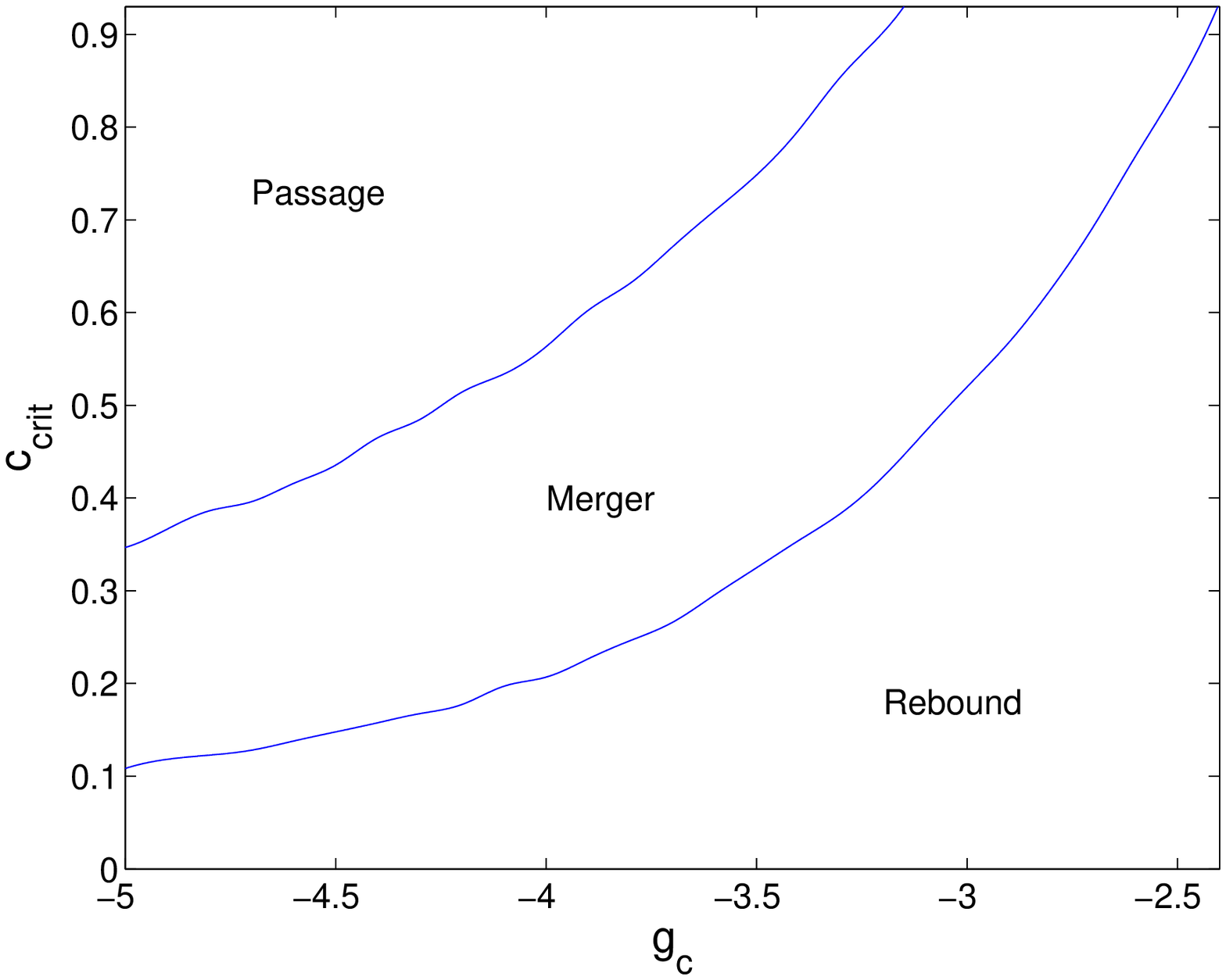} & %
\includegraphics[width=.5\textwidth]{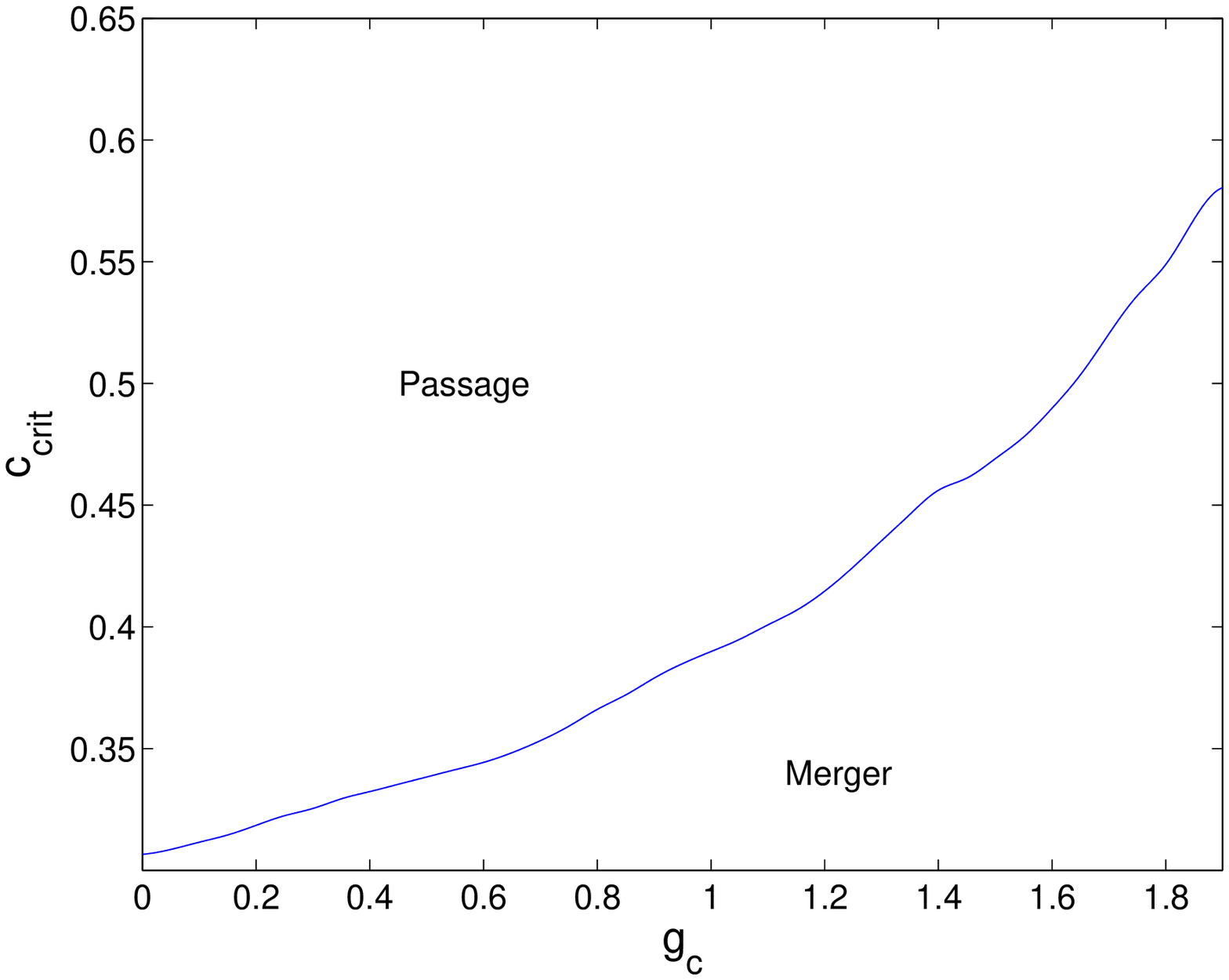} \\
(a) & (b)%
\end{tabular}%
\end{center}
\caption{(Color online) Critical values of the collision velocity versus the
strength of the contact attraction or repulsion for $g_{d}=1$ (a) and $%
g_{d}=-1$ (b). In both cases, $\protect\mu =-1$ and $\protect\epsilon =0$.}
\label{fig:collcrit}
\end{figure}

Dependences of critical values of collision velocities, which separate
different outcomes of the collision, on the strength of the local
interaction are displayed in Fig. \ref{fig:collcrit}.

In the presence of the OL, the solitons can be made mobile (by application
of a kick to them) if the nonlinearity in the model is weak enough;
otherwise, the respective Peierls--Nabarro is very high. In the
weak-nonlinearity regime, the mobility of GSs in the present model is quite
similar to that reported in the local model with the self-repulsive
interactions \cite{Sakaguchi}, as well as in the discrete model including
the DD interactions \cite{Belgrade}.

\section{Conclusion}

This work presents results of the systematic analysis of one-dimensional
bright solitons supported by contact and dipole-dipole (DD) interactions of
opposite signs in BEC. In the absence and in the presence of the optical
lattice (OL), stable soliton families have been found for the cases of local
attraction and DD repulsion or vice versa. In particular, free-space
solitons can be supported by arbitrarily weak local attraction if the DD
repulsion is fixed; in the opposite case, there is a maximum value of the
strength of the local repulsion, beyond which solitons do not exist (which
was explained in an analytical form). In the model including the OL, a
notable finding is a region of stability of subfundamental solitons (SFSs)
in the second finite bandgap. It is noteworthy too that, as seen in Figs. %
\ref{fig:existra} and \ref{fig:existar}, the gap solitons (GSs) exist, in
the case of the attractive DD interaction, if the contact repulsion is
strong enough, and, in the opposite case of the repulsive DD interaction,
GSs exist if the contact attraction is not too strong.

Collisions between bright solitons in the free space were considered too.
The collision scenario is the usual one in the case of the local attraction
(merger and quasi-elastic passage at small and large velocities,
respectively), while in the opposite case, when the local interaction is
repulsive, a region of rebound was additionally found at smallest values of
the velocities, which is explained by the long-range repulsion between the
solitons.

\section*{Acknowledgment}

B.A.M. appreciate hospitality of the Nonlinear Physics Group of the
University of Seville (Spain). The work of this author is supported, in a
part, by grant No. 149/2006 from the German-Israel Foundation. P.G.K.
gratefully acknowledges support from NSF-CARREER, NSF-DMS-0806762 and from
the Alexander von Humboldt Foundation. The work of D.J.F. was partially
supported by the Special Research Account of the University of Athens. J. C.
acknowledges financial support from the Ministerio of Ciencia e Innovaci\'on
of Spain, project number FIS2008-04848.

\end{document}